\documentclass[useAMS,usenatbib]{mnras}
\usepackage{graphicx}
\usepackage{amsmath}
\usepackage{amssymb}
\graphicspath{{./images/}}
\usepackage{amssymb,epsfig,fancyhdr,slashed}
\usepackage{pstricks,pstricks-add,pst-node,pst-text,pst-plot,pst-3d}
\usepackage{comment}
\usepackage{ulem}

\newcommand{\be}{\begin{equation}}
\newcommand{\ee}{\end{equation}}
\newcommand{\bear}{\begin{eqnarray}}
\newcommand{\ear}{\end{eqnarray}}
\newcommand{\nline}{\nonumber \\}
\newcommand{\f}{\frac}
\newcommand{\de}{{\rm d}}

\newcommand{\eqn}[1]{equation~\eqref{#1}}

\renewcommand{\mathbf}[1]{\mbox{\boldmath $#1$}}

\newcommand{\specialcell}[2][c]{%
  \begin{tabular}[#1]{@{}c@{}}#2\end{tabular}}

\newcommand{\HI}{\rm H~{\sc i }}
\newcommand{\HII}{\text{H~{\sc ii }}}

\newcommand{\TB}{\delta T_{\rm b}}
\newcommand{\MSUN}{{\rm M}_{\odot}}

\newcommand{\XHI}{x_{\rm HI}}
\newcommand{\XHII}{x_{\rm HII}}
\newcommand{\TS}{T_{\rm S}}
\newcommand{\TK}{T_{\rm K}}
\newcommand{\TCMB}{T_{\gamma}}
\newcommand{\lya}{\rm {Ly{\alpha}}}
\newcommand{\OmegaB}{\Omega_{\rm B}}
\newcommand{\Omegam}{\Omega_{\rm m}}

\title[Constraining the properties of ionized bubbles]{Bayesian approach to constraining the properties of ionized bubbles during reionization}
\author[Ghara \& Choudhury]{
\parbox[t]{\textwidth}{
Raghunath Ghara$^{1,2,3}$\thanks{Email: ghara.raghunath@gmail.com},
T. Roy Choudhury$^4$ } 
\vspace*{6pt} \\
$^{1}$The Oskar Klein Centre, Department of Astronomy, Stockholm University, AlbaNova, SE-10691 Stockholm, Sweden\\
$^{2}$Department of Natural Sciences, The Open University of Israel, 1 University Road, PO Box 808, Ra'anana 4353701, Israel \\
$^{3}$Department of Physics, Technion, Haifa 32000, Israel\\
$^4$ National Centre for Radio Astrophysics, TIFR, Post Bag 3, Ganeshkhind, Pune 411007, India\\}

\date{Accepted XXX. Received YYY; in original form ZZZ}

\pubyear{2019}

\begin{document}
\label{firstpage}
\pagerange{\pageref{firstpage}--\pageref{lastpage}}
\maketitle
\begin{abstract}
A possible way to study the reionization of cosmic hydrogen is by observing the large ionized regions (bubbles) around bright individual sources, e.g., quasars, using the redshifted 21~cm signal. It has already been shown that matched filter-based methods are not only able to detect the weak 21~cm signal from these bubbles but also aid in constraining their properties. In this work, we extend the previous studies to develop a rigorous Bayesian framework to explore the possibility of constraining the parameters that characterize the bubbles. To check the accuracy with which we can recover the bubble parameters, we apply our method on mock observations appropriate for the upcoming SKA1-low. For a region of size $\gtrsim 50$ cMpc around a typical quasar at redshift 7, we find that $\approx 20$ h of integration with SKA1-low will be able to constrain the size and location of the bubbles, as well as the difference in the neutral hydrogen fraction inside and outside the bubble, with $\lesssim 10\%$ precision. The recovery of the parameters are more precise and the SNR of the detected signal is higher when the bubble sizes are larger and their shapes are close to spherical. Our method can be useful in identifying regions in the observed field which contain large ionized regions and hence are interesting for following up with deeper integration times.
\end{abstract}




\section{Introduction}
\label{sec:intro}

Detecting and understanding the properties of the very first radiating sources in the Universe is one of the frontiers of modern astronomy. Detecting these sources directly using presently available facilities is challenging as they are apparently faint and also the observations are limited by systematic effects.  Nevertheless, a significant amount of progress has been achieved towards detecting these high redshift galaxies  \citep{Hu10, Kashikawa11, Ellis13, 2017arXiv171102090B} and quasars \citep{Fan06b, Mortlock11, Venemans15,  2018Natur.553..473B} through various surveys over last few decades. 

An indirect way to detect these sources would be through their effect on the neutral gas (mainly hydrogen and helium) in the intergalactic medium (IGM). It is believed that the radiation from these primordial sources would modify the ionization and thermal state of the neutral hydrogen (\HI) IGM during the Cosmic Dawn (CD) and the Epoch of Reionization (EoR). The signal from the cosmological \HI can, in principle, be detected through the redshifted 21~cm signal using low-frequency radio telescopes. In fact, several existing radio interferometers such as the Low Frequency Array (LOFAR)\footnote{http://www.lofar.org/} \citep{van13, 2017ApJ...838...65P, 2020MNRAS.493.1662M}, the Precision Array for Probing the Epoch of Reionization (PAPER)\footnote{http://eor.berkeley.edu/} \citep{parsons13}, the Murchison Widefield Array (MWA)\footnote{http://www.mwatelescope.org/} \citep{bowman13, tingay13} etc. have dedicated their valuable resources to detect this signal from the EoR. Highly sensitive future telescopes such as the Square Kilometre Array (SKA)\footnote{http://www.skatelescope.org/} and Hydrogen Epoch of Reionization Array (HERA)\footnote{https://reionization.org/} will also observe this signal from the Cosmic Dawn as well. 

Since the expected cosmological signal is much weaker than the system noise and the astrophysical foregrounds, one requires a good theoretical understanding of the generation of the signal for interpreting the data. Over the last few decades, several theoretical studies have been aimed towards understanding the details of the EoR. These studies include analytical calculations \citep[e.g.,][]{furlanetto04, 2014MNRAS.442.1470P, 2019JCAP...04..051N, 2020MNRAS.492..634G}, semi-numerical simulations \citep{zahn2007, mesinger07, santos08, Thom09, choudhury09,  2020MNRAS.493.4728G,  2018MNRAS.481.3821C}, and full radiative transfer numerical simulations \citep{Iliev2006,  mellema06, McQuinn2007, shin2008, baek09}. The general consensus from these studies is that the ultra-violet (UV) photons from the first sources create ionized bubbles which later overlap and complete the reionization of primordial \HI around redshift $\approx 5.5 - 6$ \citep{Fan06b, 2011ApJ...728L...2S, 2018arXiv180706209P, 2019MNRAS.485L..24K}. Using these theoretical models, the detectability of the signal has been extensively studied in terms of various statistical quantities such as the power spectrum \citep{2015MNRAS.449.4246G, 2017MNRAS.468.3869S, 2018MNRAS.475.1213S}, the bi-spectrum \citep{2018MNRAS.476.4007M, 2019JCAP...02..058G} and the size distribution of the ionized bubbles \citep{2018arXiv180106550G}. All these studies show that the signal detectability critically depends on the population and properties of these early sources. 

A method complimentary to using the statistical quantities is to detect the signature of ionized bubbles around the individual sources, e.g., using the \HI 21~cm maps \citep{2015aska.confE..10M, ghara16}. In fact, the characteristic sizes of these bubbles are expected to be as large as several tens of comoving megaparsec (cMpc) at a stage when the universe is only 50$\%$ ionized \citep[see e.g,][]{2018arXiv180106550G}. Nevertheless, since the strength of the 21~cm signal is weak, the detection of such large \HII bubbles still remains an open challenge.

A possible method to detect these individual bubbles is by using the matched filter \citep{kanan2007MNRAS.382..809D}. The technique applies predefined filters, such as spherical top-hat filters in the image space, to the measured visibilities for enhancing the detectability of \HII regions from the EoR. Previous studies such as \citet{kanan2007MNRAS.382..809D, 2012MNRAS.426.3178M} have shown that this is an efficient method to detect the large isolated \HII bubbles during the EoR. For example, they found that \HII regions with size $\gtrsim 25$ cMpc should be easily detected for $\sim1200$ h of observation with LOFAR \citep{datta2012a}. The method has been extended also to the Cosmic Dawn by \citet{ghara15c} where one expects isolated regions with strong absorption signal around the individual/clustered sources \citep[see e.g.,][]{ghara15a}.

The detection of luminous quasars as early as at redshift $\gtrsim 7$ \citep{Mortlock11, 2018Natur.553..473B} suggests that the size of \HII regions around those rare sources will be larger compared to those around the galaxies hosting stars. Thus, the locations of these quasars are suitable for targetted searches using matched filtering techniques. In fact, \citet{2012MNRAS.426.3178M} have shown that it is, in principle, possible to constrain the properties of the quasars and the surrounding IGM through the matched filtering method with present telescopes, although the observational time required can be quite large. Nevertheless, this technique can accurately estimate the age of the quasar at the center if we have prior knowledge on the emission rate of the ionizing photons from the observation of Infrared spectrum, and can potentially place a lower limit on the neutral fraction at that redshift \citep[][]{2012MNRAS.426.3178M}.  The study also assumed that the observations will be targetted towards the known quasars. In general, however, such favourable situations may not be common during the EoR. Thus, it is important to develop a framework to detect \HII regions through a blind search in 21~cm maps where no prior information about the locations of the ionized regions is known. In addition, it is also important to develop methods to constrain the properties of the bubbles (e.g., the position and the size) through the same signal. Such detection of the large \HII regions using the observations of 21~cm signal from the EoR will also imply indirect detection of the 21~cm signal in general. 

In this work, our goal is to develop a mathematically rigorous method to detect and measure the quantities that characterize the properties of the bubble. We essentially extend the method used in studies such as \citet{kanan2007MNRAS.382..809D,2012MNRAS.426.3178M} to a Bayesian approach with the aim of constraining the properties of the ionized bubbles. To achieve this, we make use of methods developed for detecting the weak gravitational waves from isolated sources buried under the detector noise \citep{1992PhRvD..46.5236F}. The method allows one to calculate the likelihood of detecting a bubble buried within the telescope noise. The likelihood thus defined depends on the model used for the signal which, in turn, depends on several parameters (e.g., the position and size of the bubble along with the neutral hydrogen fraction of the surrounding medium). One can implement the already existing and widely-used Markov Chain Monte Carlo (MCMC) algorithms to explore the multi-dimensional parameter space of the filter and obtain the corresponding posterior probability distribution. We apply this framework to simulated mock visibilities of SKA1-low which contain the expected signal and system noise. This allows us to check how accurately do the inferred parameter values match the input signal and also to work out the expected precision with which the parameters can be recovered.

This paper is structured in the following way. In section \ref{sec:framework}, we describe the basic methodology including the likelihood calculations used in this study.  In section \ref{sec:sim}, we describe the mock visibilities used in this paper. In particular, the different subsections describe simulations of the 21~cm signal and the calculation of system noise used in this study. We present our results in section \ref{sec:res} before we conclude in section \ref{sec:con}.  The cosmological parameters used in this study are  $\Omegam=0.32$, $\Omega_\Lambda=0.68$, $\OmegaB=0.049$, $h=0.67$, $n_{\rm s}=0.96$, and $\sigma_8=0.83$, which are consistent with $Planck$ \citep{Planck2013}. 

\section{Framework}
\label{sec:framework}

The directly measurable quantities in a radio interferometric observation are the visibilities. The measured visibilities $V(\vec{U}, \nu)$ depend on the baseline vector $\vec{U}$ and the observation frequency $\nu$. The baseline for a pair of individual antennas of the interferometer is given by $\vec{U}=\vec{d}/\lambda$, where  $\vec{d}$ is the distance between the two antennas and $\lambda = c / \nu$ is the observing wavelength. Besides the 21~cm signal, the measured visibilities will also contain foregrounds from our Galaxy and extragalactic point sources, the instrumental noise, and other systematic effects (e.g., the ionosphere and radio frequency interference). In this study,  we assume that the foregrounds and other systematic effects can be identified and subtracted perfectly from the visibilities.  Under such an assumption, the visibilities contain only the redshifted 21~cm signal and system noise. Thus, we can write  
\begin{equation}
V(\vec{U}, \nu) = S(\vec{U}, \nu) + N(\vec{U}, \nu),
\label{eq:visi}
\end{equation}
where $S$ and $N$ represent the contributions from the signal and system noise respectively. 

The main principle of this framework is to smooth the measured visibilities using an appropriate filter and define a likelihood which can be used for Bayesian parameter estimation. In the matched filter formalism, the smoothing filter is chosen as whatever we expect the signal to be. In this sense, the filter would depend on the parameters required to model the signal.

Unfortunately, the characterization of the expected signal in terms of a few parameters is not straightforward as the signal is product of several complex physical processes. Due to this, earlier studies like \citet{kanan2007MNRAS.382..809D} have used simple filters which assumes spherical \HII regions around the sources. We too follow the same approach and use the simple spherical top-hat filter in this work. It is important to keep in mind that this may not necessarily be the optimum filter for detecting \HII bubbles as their shapes in the map may be more complex because of significant overlap between individual \HII regions. However, as we will show later, this simple-minded filter serves our purpose adequately for measuring the position and size of the bubble.

We characterize the spherical top-hat filter in terms of five parameters as given below.
\begin{itemize}
\item $R$ (cMpc): radius of the spherical top-hat filter. 
\item $\theta_X$ (arcmin): $x$-coordinate of the centre of the filter along the angular direction.
\item $\theta_Y$ (arcmin): $y$-coordinate of the centre of the filter along the angular direction.
\item $\Delta\nu$ (MHz): frequency difference between the central frequency channel of the radio observation and the channel that contains the centre of the filter.
\item $A_{\TB}$ (mK): the amplitude of the signal outside the \HII bubble. We set the signal to zero inside the bubble.
\end{itemize}

We label the set of these five parameters as $\mathbf{\mu} \equiv \left\{R,\theta_X,\theta_Y,\Delta\nu, A_{\TB} \right\}$. Let us denote the Fourier transform of the top-hat filter, appropriate for the visibility space, as $S_f(\vec{U}, \nu; \mathbf{\mu})$.

\subsection{Likelihood calculation}
\label{sec:likelihood}

In a Bayesian framework, the conditional probability that a particular signal $S_f(\vec{U}, \nu; \mathbf{\mu})$, characterized by the parameters $\mathbf{\mu}$, is present in the data $V(\vec{U}, \nu)$ is proportional to  \citep{1992PhRvD..46.5236F}
\begin{align}
  \Lambda(\mathbf{\mu}) &= p(\mathbf{\mu})~\exp\left\{\f{1}{\sigma_{\rm rms}^2} \int \de^2 U \int \de \nu~\rho_B(\vec{U}, \nu) \right.
                          \nline
                          & \left. \times \left[ 2 V(\vec{U}, \nu) S_f^*(\vec{U}, \nu; \mathbf{\mu}) - |S_f(\vec{U}, \nu; \mathbf{\mu})|^2 \right] \right\},
\end{align}
where $p(\mathbf{\mu})$ is a prior probability that the signal $S_f(\vec{U}, \nu; \mathbf{\mu})$ can be characterized by the parameters $\mathbf{\mu}$. As described in the previous section, we aim to detect large \HII regions in this study. Thus, the particular signal $S_f(\vec{U}, \nu; \mathbf{\mu})$ considered here is the Fourier transform of the top-hat filter (or a spherical ionised region in our case) characterised by five parameters as $\mathbf{\mu} \equiv \left\{R,\theta_X,\theta_Y,\Delta\nu, A_{\TB} \right\}$. The quantity $\rho_B(\vec{U},\nu)$ is the normalized baseline distribution which satisfies the condition 
\begin{equation}
    \int \de^2U \int \de \nu ~\rho_B(\vec{U},\nu) = 1,
  \end{equation}
  and $\sigma_{\rm rms}$ is the rms of the system noise map. The details of this calculation are given in Appendix \ref{appen:like}. We should point out that this conditional probability is nothing but the posterior distribution of the parameters and hence can be used to obtain the confidence intervals on the multi-dimensional parameter space using conventional Bayesian parameter estimation methods.

For a uniform prior, we can write the logarithm of the likelihood as
\begin{align}
  \log \Lambda(\mathbf{\mu}) & = \frac{1}{\sigma_{\rm rms}^2} \int {\rm d}^2 U \int {\rm d}\nu  ~\rho_B(\vec{U},\nu)
  \nonumber\\  
&\times \left[2 V(\vec{U},\nu) ~S_f^{*}(\vec{U}, \nu; \mathbf{\mu}) - |S_f(\vec{U},\nu; \mathbf{\mu})|^2\right].
\label{equ_exp_like}
\end{align}
Given the data $V$, the likelihood can be computed for any parameter set $\mathbf{\mu}$. The best-fit parameter $\mathbf{\hat{\mu}}$ is the one that maximizes the likelihood. As already mentioned, the way the likelihood is defined above, one can also obtain the confidence contours using standard Bayesian methods. In this study, we use the Markov Chain Monte Carlo-based publicly available code {\sc cosmomc}\footnote{\tt https://cosmologist.info/cosmomc/} \citep{2002PhRvD..66j3511L} to explore the five-dimensional parameter space of the model signal and to estimate the best fit parameter value $\mathbf{\hat{\mu}}$ and the confidence intervals.

We also note that a related quantity is the signal-to-noise ratio defined as \citep{1992PhRvD..46.5236F},
\begin{equation}
\mbox{SNR} = \left[\f{1}{\sigma_{\rm rms}^2} \int \de^2 U \int \de \nu~\rho_B(\vec{U}, \nu)~|S_f(\vec{U}, \nu; \mathbf{\hat{\mu}})|^2 \right]^{1/2},
\label{snr_bestfit}
\end{equation}
which essentially measures the strength of detected signal.

\section{Mock visibilities}
\label{sec:sim}

In order to test how well the likelihood-based analysis recovers the underlying true properties of the ionized bubbles with the upcoming telescopes, we need to generate mock visibilities appropriate for these instruments. In this work, we focus on visibilities appropriate for the upcoming SKA1-low. For this, we consider the presently announced antenna configuration of SKA1-low. This revised configuration of the SKA1-low consists of 512 antenna stations\footnote{\url{https://astronomers.skatelescope.org/wp-content/uploads/2016/09/SKA-TEL-SKO-0000422_02_SKA1_LowConfigurationCoordinates-1.pdf}}. In our mock observation, we assume that a region with Right Ascension $0^\circ$ and declination $-30^\circ$ is observed for 4 hours daily. The details of the mock observation are given in Table \ref{table_obs}.

\begin{table}
\centering
\small
\tabcolsep 2pt
\renewcommand\arraystretch{1.5}
 \begin{tabular}{c c c}
\hline
Number of antenna ($N_{\rm ant}$) & 512 	 \\
Effective area ($A_{\rm eff}$) & 962 m$^2$\\
Integration time ($\Delta t_c$) & 120 sec \\
Observation time per day & 4 h \\
Total observation time ($t_{\rm obs}$) & 20 h \\
Declination to observed region & -30$^\circ$ \\
Right Ascension & 0$^\circ$ \\
Observation frequency ($\nu_c$) & 177.5 MHz \\
Band width ($B_\nu$) & 18.94 MHz \\
Frequency resolution ($\Delta \nu$) & 175 kHz \\
System temperature ($T_{\rm sys}$)  & $\left[100 + 60\times(\frac{\nu}{300\mbox{MHz}})^{-2.55}\right]$~K \\
\hline
\end{tabular}
\caption[]{The details of the mock observation with SKA1-low. The frequency of observation is chosen to observe the 21~cm signal at redshift 7. The baseline $uv$ coverage corresponds to an integration time of 120 seconds and 4 hours of observation per day, while the total observation time is 20 hours.}
\label{table_obs}
\end{table}

As mentioned in \eqn{eq:visi}, the measured visibility is the sum of the cosmological signal and the system noise (ignoring the foregrounds and other systematics). The methods used to generate the signal and noise are described next.

\begin{figure*}
\begin{center}
 \includegraphics[width=0.9\textwidth]{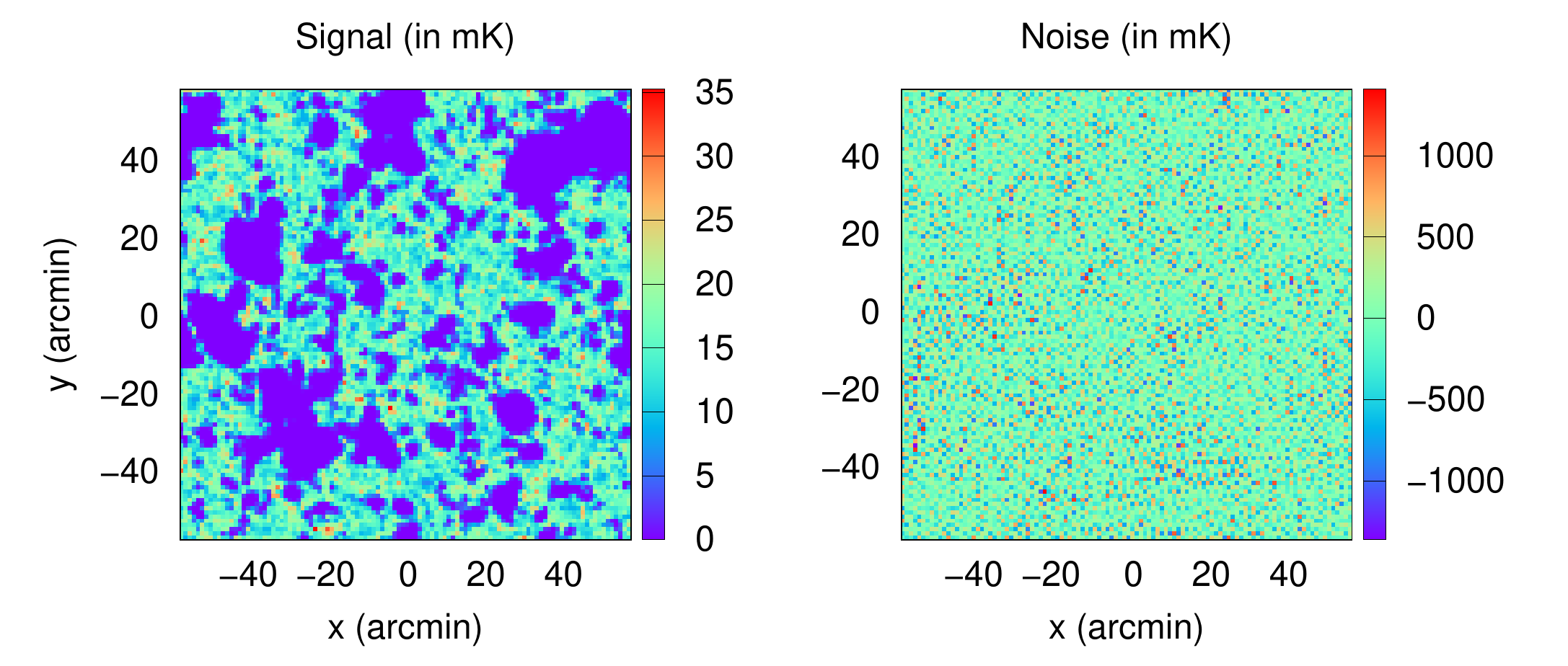}
 \caption{\textit{Left panel}: A two-dimensional slice of the brightness temperature from the simulated light-cone. This slice is chosen such that it contains the halo with the largest mass at redshift $\approx$7. The resolution of the map is $1.075'$ which corresponds to a length scale of $\approx 2.76$ cMpc. \textit{Right panel}: The system noise map with $\sigma_{\rm rms} =  35.3 ~\mu$Jy per beam (370 mK). This corresponds to 20 h of observation with SKA1-low with a frequency channel width of 175 kHz.}
   \label{fig:images1}
\end{center}
\end{figure*}

\subsection{The cosmological \HI 21~cm signal}
\label{sec:sig}
The expected 21~cm signal is simulated using the code {\sc grizzly} \citep{ghara15a, ghara18} which uses one-dimensional radiative transfer schemes to generate the brightness temperature cubes from the outputs of an $N$-body simulation. Details of the $N$-body simulation as well as the method are given below.

\subsubsection{$N$-body simulation}
\label{n_body}
The dark matter only $N$-body simulation was carried out using code {\sc cubep}$^3${\sc m}\footnote{\tt http://wiki.cita.utoronto.ca/mediawiki/index.php/CubePM} \citep{Harnois12}. The simulation provides snapshots of density and velocity fields and also the list of haloes from redshift 20.134 to redshift 6. The time difference between two successive snapshots  is 10 Myr. The grid dimension of the density and velocity fields is $216^3$, while the size of the simulation box is $200/h$ cMpc. The minimum mass of the haloes identified using the spherical over-density method \citep[see e.g.,][]{2013MNRAS.433.1230W} is $2.2\times 10^9 ~\MSUN$. The same $N$-body simulation was previously used in \citet{ghara15b}.

\subsubsection{Modelling the radiation sources}
\label{sec:source}

We assume that each halo identified from the $N$-body simulation contains a galaxy. The radiation from the galaxy is contributed by two components: stars and a mini-quasar powered by the central black hole. The stellar population is taken to be the Population II stars which turn out to be the main contributors to the hydrogen ionizing photons. We assume the stellar mass of the galaxy is proportional to the mass of the host dark matter halo with the proportionality constant fixed such that the number of ionizing photons emitted per second per stellar mass in the dark matter halo is $\approx 5.67\times 10^{43}$. This choice gives rise to reionization history such that the reionization ends around redshift 6.  The spectral energy distribution (SED) of the stars is generated using the publicly available code {\sc pegase} \citep{Fioc97, ghara15a}.

For the mini-quasars, the SED is taken to be a simple power-law with spectral index 1.5. The X-ray luminosity\footnote{We assume that the X-ray band spans from 100 eV to 10 keV, while the UV range is 13.6-100 eV.}, produced predominantly by the mini-quasars, is taken to be 5$\%$ of the UV luminosity of the galaxy. With these assumptions, the spectrum of the galaxy is entirely determined by the host dark matter halo mass.

\begin{figure*}
\begin{center}
 \includegraphics[width=0.7\textwidth]{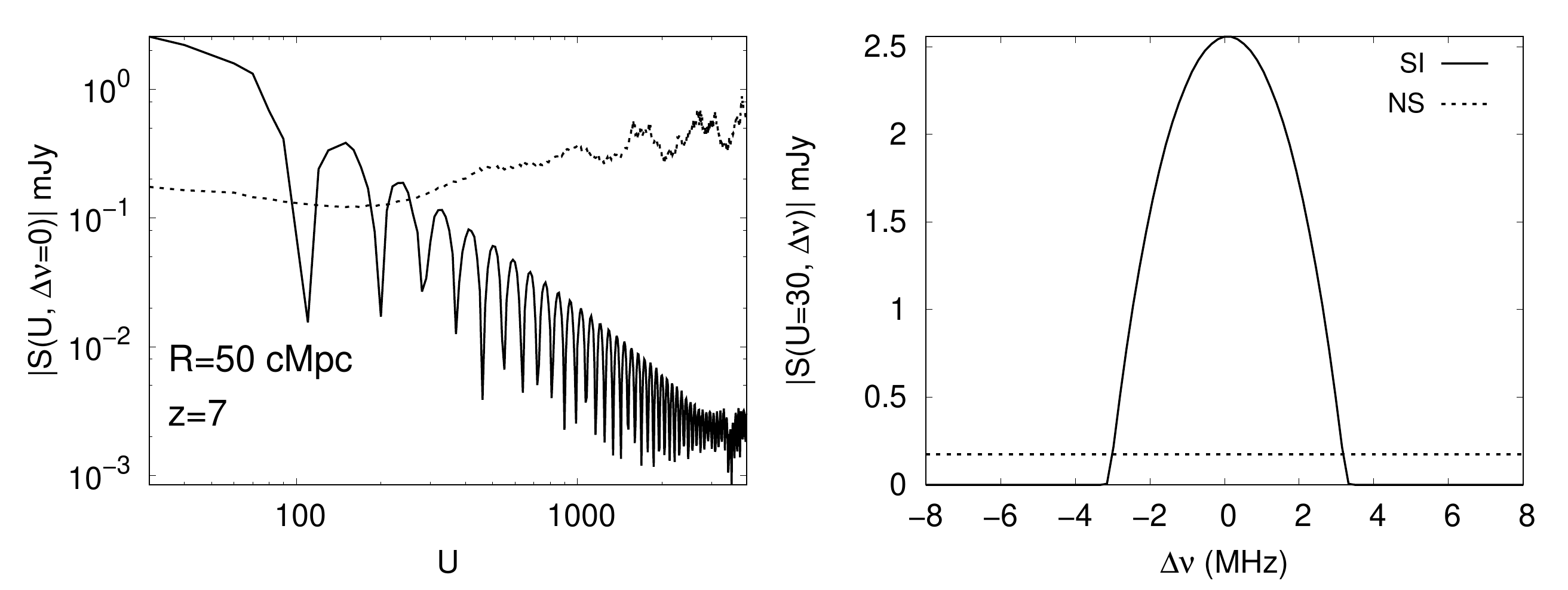}
 \caption{\textit{Left panel}: The amplitude of the visibilities as a function of baseline length $U$ for an isolated \HII bubble of size 50 cMpc placed in a uniform neutral medium at redshift 7. The bubble centre coincides with the centre of the field of view. The frequency channel corresponds to the one which contains the centre of the test bubble at redshift 7. The dotted line shows the noise rms for 20 h observation with SKA1-low with the other observation parameters given in Table \ref{table_obs}.
\textit{Right panel}: The amplitude of the signal visibilities as a function of $\Delta \nu=\nu-\nu_c$, the frequency deviation from the central channel $\nu_c$. The baseline length is chosen to be $U = 30$ for this plot. The dotted line shows the noise rms for 20 h observation with SKA1-low with the other observation parameters given in Table \ref{table_obs}.}
   \label{fig:visi}
\end{center}
\end{figure*}

\subsubsection{Simulation of \HI maps using {\sc grizzly}}
\label{sec:sim_grizzly}

We begin by creating a library which consists of a large number of one-dimensional profiles of $\XHII$, $\TK$ around isolated sources for different combinations of dark matter halo masses, redshifts and density contrast. These profiles are used to generate the ionization maps and kinetic temperature maps in {\sc grizzly}. In this code, we first estimate the size of ionized regions around each source and assign \HII regions around the sources using the library already created. At the same time, we also compute the ``unused'' ionizing photons in the overlapped regions. These unused photons are then redistributed among the overlapping sources by increasing the size of the ionized regions appropriately. The ionization fraction in the partially ionized regions is calculated using a simple overlap prescription and the kinetic temperature maps are generated using a correlation between the ionization fraction and $\TK$ \citep[for details, see][]{ghara15a, 2019MNRAS.487.2785I}. Finally, we generate the $\lya$ flux maps assuming that the $\lya$ photons flux decrease as $1/R^2$ where $R$ is the radial distance from the source. We generate the spin temperature $T_{\rm S}$ maps using this $\lya$ photon field assuming the collisional coupling is negligible at the redshift of our interest.

It is straightforward to generate the brightness temperature maps from the maps of density contrast ($\delta$), neutral fraction ($\XHI$) and spin temperature ($\TS$) using \citep[see e.g.,][]{madau1997, Furlanetto2006}
\begin{align}
 \TB (\vec{\theta}, \nu)  & = 27 ~ x_{\rm HI} (\mathbf{x}, z) [1+\delta_{\rm B}(\mathbf{x}, z)] \left(\frac{\OmegaB h^2}{0.023}\right) \nonumber\\
&\times \left(\frac{0.15}{\Omegam h^2}\frac{1+z}{10}\right)^{1/2}\left[1-\frac{\TCMB(z)}{T_{\rm S}(\mathbf{x}, z)}\right]\,\rm{mK},
\label{brightnessT}
\end{align}
where $\vec{\theta}$ is the angular position of the observed region and $\nu=1420/(1+z)$ MHz is the frequency of observation. $\TCMB(z)$ = 2.73 $\times (1+z)$ K denotes the  brightness temperature of the Cosmic microwave background (CMB). It is straightforward to convert the three-dimensional position $\mathbf{x}$ in the simulation box to an angle $\vec{\theta}$ on the sky and a frequency $\nu$ along the line of sight.

While generating the $\TB$ maps, we also include the effect of the peculiar velocities of the gas in the IGM by moving the gas elements appropriately along the line of sight. In addition, we include another line of sight effect, popularly known as the ``light-cone'' effect which accounts for the evolution of the signal with redshift or observational frequency \citep{ghara15b}. Finally, we reduce the resolution of the simulated map by factor 2 using a top-hat smoothing filter in order to reduce the computation time of our analysis. Our final two-dimensional maps thus have $108^2$ grids.

 The left-hand panel of Figure \ref{fig:images1} presents one such two-dimensional slice from the light-cone at redshift $\approx$7. The angular resolution of the map is $1.075'$ which corresponds to 2.76 cMpc length on the sky plane. The volume averaged ionization fraction at redshift 7 is $\approx 0.4$ for the chosen reionization model. The kinetic temperature of the gas in the IGM is well above the CMB brightness temperature due to significant X-ray heating. This fact, with a strong $\lya$ coupling, makes the signal almost insensitive to the spin temperature fluctuations at that redshift. Thus, the fluctuation in $\TB$, as shown in Figure \ref{fig:images1}, is solely due to the fluctuation in ionization fraction and density field, see \eqn{brightnessT}.

The gridded visibilities of the signal are generated by performing a discrete Fourier transform of the two-dimensional slice at each frequency channel (an example of such a channel is shown in the left-hand panel of Figure \ref{fig:images1}):
\begin{equation}
S(\vec{U}, \nu) = F(\vec{U}, \nu) \int d^2\theta ~I_{\rm S}(\vec{\theta}, \nu)~A(\vec{\theta})~e^{i2\pi \vec{\theta} \cdot \vec{U}},
\label{visibi}
\end{equation}
where the sky specific intensity at frequency $\nu$ can be related to $\TB$ as,
\begin{equation}
I_{\rm S}(\vec{\theta},\nu) = \frac{2 k_{\rm B} \nu^2  }{c^2} \TB (\vec{\theta}, \nu).
\label{inten}
\end{equation}
Here, $A(\vec{\theta})$ denotes the primary beam pattern of individual antenna. We have assumed a rather wide primary beam which allows us to set $A(\vec{\theta}) \approx 1$ for the rest of the work. The quantity $F(\vec{U}, \nu)$ represents the sampling function of the baselines, i.e., $F = 1$ for the grid points with at least one measured visibility, and $F = 0$ otherwise. 
Note that $S(\vec{U}, \nu)$ denotes the observed visibilities of the signal in general while  the quantity $S_f(\vec{U}, \nu; \mathbf{\mu})$ in Section \ref{sec:framework} denotes the top-hat smoothing filter used in this study.

\begin{table*}
\centering
\small
\tabcolsep 6pt
\renewcommand\arraystretch{1.5}
   \begin{tabular}{c c c c c}
\hline
\hline
Parameters & Fixed value & Explored range & Best fit  & Standard deviation 	 \\

\hline
\hline
$R$ (cMpc) & 50.0   & 10.0, 100.0   &  50.11      & 0.4	  			\\
$\theta_X$ (arcmin) & 0.0 & -40.0, 40.0	& 0.019	&	0.32	\\
$\theta_Y$ (arcmin) & 0.0	& -40.0, 40.0 & 0.017	&	0.3 \\
$\Delta \nu$ (MHz) & 0.0	& -4.0, 4.0 & -0.002	&	0.03 \\
$A_{\TB} (\rm mK)$  & 30.0  & -100, 100  & 29.65 & 1.3\\
\hline
\hline
\end{tabular}
\caption[]{This shows the outcomes of an MCMC analysis. $\theta_X, \theta_Y$ are the angular positions of the centre of the filter of radius $R$. $\Delta \nu=\nu-\nu_c$ is the frequency deviation from the central channel $\nu_c$ and $A_{\TB}$ is the amplitude of the filter signal. The second column shows the chosen values of these parameters for the test \HII bubble as our input signal. The best fit values of the parameters are obtained from an MCMC analysis and the standard deviations correspond to 20 h of observation with SKA1-low with a bandwidth of 18.96 MHz at redshift 7. The details of the mock observation are given in Table \ref{table_obs}.  }
\label{tab:testbub}
\end{table*}

\subsection{Noise simulation}
\label{ns_sim}

We assume that the system noise $N(\vec{U}, \nu)$ at different baselines and frequency channels are uncorrelated. The rms noise for each  baseline, polarization and frequency channel can be written as
\begin{equation}
\sigma_N = \frac{\sqrt{2} k_B T_{\rm sys}}{A_{\rm eff} \sqrt{\Delta \nu_{c} ~ \Delta t_{c}}},
\end{equation}
where $\Delta \nu_c$ and $\Delta t_{c}$ are the frequency channel width and the correlator integration time respectively. The quantities  $A_{\rm eff}$ and $T_{\rm sys}$ represent the effective collecting area of individual antenna and the system temperature respectively. The values of these quantities as used in this study can be found in Table \ref{table_obs}. The noise at each baseline and frequency channel can be reduced by integrating over a longer time $t_{\rm obs}$, in which case we can scale the rms by $\sqrt{\Delta t_c / t_{\rm obs}}$ and obtain
\be
\sigma_N = \frac{\sqrt{2} k_B T_{\rm sys}}{A_{\rm eff} \sqrt{\Delta \nu_{c} ~ t_{\rm obs}}},
\label{rms_noise}
\end{equation}

The two-dimensional maps of the noise visibilities are generated by assuming the noise to be a gaussian random variable with zero mean and rms as given in equation \ref{rms_noise}. The real space noise maps can be generated by performing Fourier transforms of those visibility maps. The rms of the noise maps thus generated can be expressed as 
\begin{align}
  \sigma_{\rm rms} &= \f{\sigma_N}{\sqrt{N_B~N_c}}
                     \nline
  & = \frac{\sqrt{2} k_B T_{\rm sys}}{A_{\rm eff} \sqrt{B_\nu ~ t_{\rm obs}N_{\rm ant}(N_{\rm ant}-1)/2}},
\end{align}
where $N_B = N_{\rm ant} (N_{\rm ant} - 1) / 2$ is the number of baselines for $N_{\rm ant}$ number of antennas in the interferometer and $N_C = B_{\nu} / \Delta \nu_c$ is the total number of channels given the total bandwidth $B_\nu$. The values of these quantities too can be found in Table \ref{table_obs}. The bandwidth used in this study corresponds to the size of our simulation box which is $200/h$ cMpc. On the other hand, the frequency resolution 175 kHz and the angular resolution $1.075'$ correspond to same length scale 2.76 cMpc, the grid resolution of the maps. The right-hand panel of Figure \ref{fig:images1} shows the noise map corresponding to the frequency channel at redshift 7 for 20 h of observation time with SKA1-low. The rms of the map is 35.3 $\mu$Jy per beam which is equivalent to a brightness temperature of 370 mK at this resolution. One can see that the rms of the noise map is quite large compared to the signal strength as shown in the left-hand panel of the same figure. The rms can be reduced further by integrating over a longer observation time and lowering the resolution \citep{ghara16}.

\begin{figure*}
\begin{center}
 \includegraphics[width=0.8\textwidth]{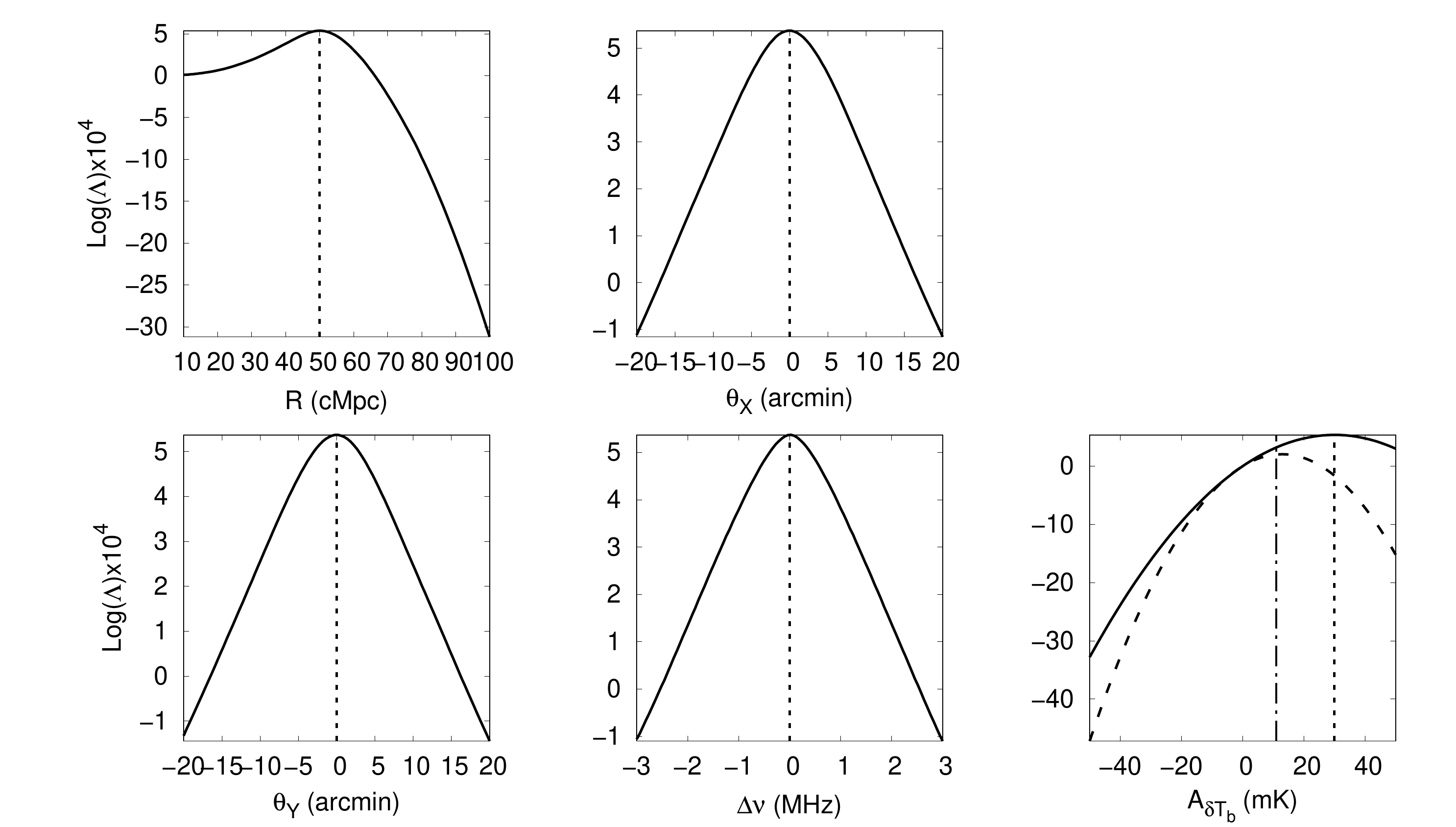}
 \caption{Different panels represent the log of the likelihood $\Lambda(\mathbf{\mu})$ as a function of five filter parameters $\mathbf{\mu}$, i.e, the radius of the sphere ($R$), angular positions of the centre of the filter ($\theta_X, \theta_Y$), position of the filter along the frequency axis ($\Delta \nu$) and the amplitude of the filter ($A_{\TB}$). Note that we vary one parameter at a time for this plot while we set the other parameters to their known values (solid curves). The input values of the parameters for this test \HII region are denoted by the dotted vertical lines.  These likelihoods correspond to a 20 hours observation time and 18.96 MHz bandwidth using SKA1-low at a central frequency 177.5 MHz. In the rightmost panel in the bottom row, the dashed curve is similar to the solid curve, but with a filter radius $R=70$ cMpc which is different than our fiducial choice. The vertical dot-dashed curve represents the $A_{\TB}$ value that corresponds to the maximum of the likelihood in this case.}
   \label{fig:mcmctest}
\end{center}
\end{figure*}

\begin{figure}
\begin{center}
 \includegraphics[width=0.4\textwidth]{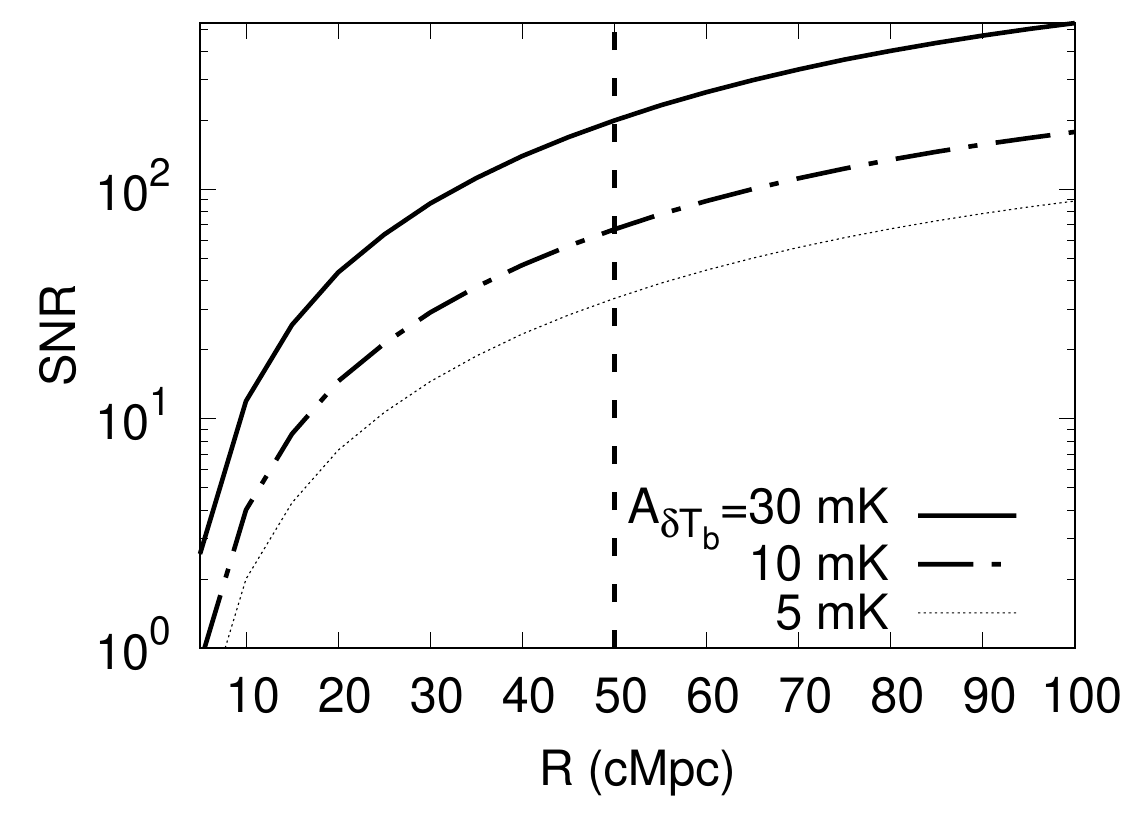}
 \caption{The Signal-to-noise ratio (SNR) as a function of the \HII bubble radius for the toy model of Section \ref{res:iso_HII}. These SNRs correspond to a 20 hours observation with 18.96 MHz bandwidth using SKA1-low at a central frequency 177.5 MHz. Different curves represent different choice of average brightness temperature $A_{\TB}$ of the signal from the IGM. The vertical dotted line corresponds to the default size of the \HII bubble used in section \ref{res:iso_HII}.}
   \label{fig:snr}
\end{center}
\end{figure}

\section{Results}
\label{sec:res}

\begin{figure*}
\begin{center}
\includegraphics[width=0.7\textwidth]{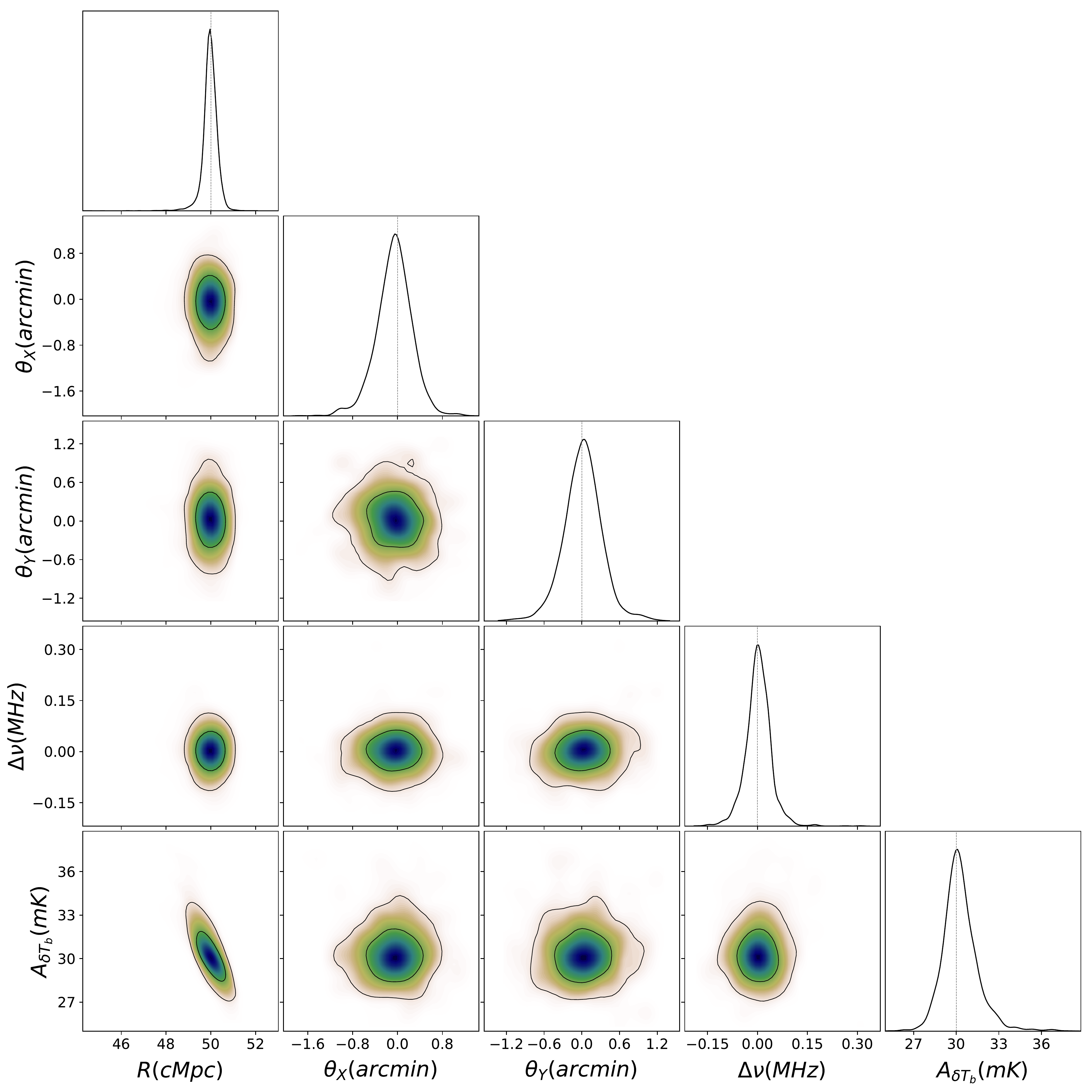}
    \caption{Posterior constraints on the \HII bubble parameters in the toy model of Section \ref{res:iso_HII} obtained through a MCMC-based analysis over the parameter space as listed in Table \ref{tab:testbub}. The contour levels in the two-dimensional contour plots represent 1$\sigma$ and 2$\sigma$ confidence levels respectively. The diagonal panels represent the marginalized probability distribution of each parameter. The vertical dotted lines represent the input values for the parameters. This analysis is done for 20 h of observation with SKA1-low at redshift 7.}
    \label{fig:images4}
\end{center}
\end{figure*}

\subsection{Toy model: an isolated {\HII} bubble in a uniform neutral medium}
\label{res:iso_HII}

Before applying our method of detecting bubbles on mock data sets generated from realistic reionization maps, we test the performance of this on a simple toy model which is that of an isolated spherical \HII bubble in a uniform neutral medium. We assume that  the centre of this \HII bubble coincides with the centre of the field of view and also with the central frequency channel (which corresponds to redshift 7).  The amplitude of the brightness temperature of the neutral medium outside the bubble is set to 30 mK, which is typical of the IGM at these redshifts. We list the values of these input parameters in the second column of Table \ref{tab:testbub}. Note that the noise map which is shown in the right panel of Figure \ref{fig:images1} will completely dominate over this toy model signal. However, such a signal can still be detected for a coarser resolution and longer observation time \citep[see e.g.,][]{ghara16}.  

The visibility amplitude of the \HII bubble at the central frequency channel as a function of the baseline length $U \equiv |\vec{U}|$ is shown in the left-hand panel of Figure \ref{fig:visi}. The visibility amplitude $|S|$ decays at the larger baselines. The oscillatory feature of this curve corresponds to the Fourier transform of the spherical \HII bubble which has sharp edges. In the right-hand panel of Figure \ref{fig:visi}, we show $|S|$ as a function of the frequency (with respect to the central channel) for a baseline $U = 30$. As expected, the amplitude decreases at frequency channels away from the centre of the test bubble. These are consistent with earlier studies such as \citet{kanan2007MNRAS.382..809D}.

The noise amplitude, generated according to the parameters given in Table \ref{table_obs}, is shown by dotted lines in both the panels of Figure \ref{fig:visi}. The number density of small baselines ($U\sim 100$) is larger than that of large baselines ($U\sim 1000$) for the proposed antenna configuration of SKA1-low \citep[see e.g.,][]{ghara15c}. This makes the noise at small baselines less compared to the large baselines. Clearly, the signal from this test \HII bubble is above the noise level and hence detectable at baselines $U\lesssim 300$ which corresponds to an angular scale of $\approx 11'$ (and a linear scale $\approx 30$~cMpc).

In Figure \ref{fig:mcmctest}, we plot the likelihood $\Lambda(\mathbf{\mu})$ as a function of all the different filter parameters for this toy model of \HII bubble. For these plots, we vary one parameter at a time while the other four parameters are fixed to their known values. One can easily see that the likelihood becomes maximum when the parameter values approach the input values as shown by the vertical dashed lines. These results thus confirm that the likelihood defined this way is a good choice for the analysis to estimate parameters.

In order to explore the effect of the filter radius on the recovery of the signal amplitude $A_{\TB}$ outside the bubble, we calculate the dependence of the likelihood on $A_{\TB}$ for $R=70$ cMpc (keeping in mind that the bubble radius is only 50~cMpc). This result is shown by the dashed curve in the rightmost panel in the bottom row of Figure \ref{fig:mcmctest}. We find that the likelihood peaks at $A_{\TB} \approx 11$ mK (the vertical dot-dashed line), which is smaller than the input value of $30$~mK. The reason for this is that the filter size is taken to be larger than the test \HII bubble size and hence, the volume of the box that is contained within the top hat filter radius contains a contribution from the neutral regions in addition to the test bubble. This reduces the contrast between the signal amplitude inside the volume of the box within the filter radius and outside. The maximum likelihood occurs when this contrast matches with the value of $A_{\TB}$. This should occur when $A_{\TB}$ is close to $({\TB}_{,\rm box}-{\TB}_{,\rm fil})$ where ${\TB}_{,\rm box}, {\TB}_{, \rm fil}$ are the mean $\TB$ for the entire box and volume of the box within the filter radius respectively. This in turn implies that the value of $A_{\TB}$ that maximizes the likelihood for a given $R$ depends on ${\TB}_{,\rm box}$ and an effective size of the ionized bubble $R_{\rm eff,HII}$ within $R$. If $[\theta_X,\theta_Y,\Delta \nu]$ are the coordinates of the centre of a spherical region of radius $R$,  we can define the effective radius $R_{\rm eff,HII}(R,\theta_X,\theta_Y,\Delta \nu)$ such that the total volume of ionized regions within the spherical region is $\frac{4\pi}{3}\times R^3_{\rm eff,HII}$, i.e., $R_{\rm eff,HII}\equiv (1-{\TB}_{, \rm fil}/{\TB}_{, \rm box})^{1/3}\times R$. The $A_{\TB}$ value that maximizes the likelihood for a $R$ is thus ${\TB}_{,\rm box}\times \frac{R^3_{\rm eff,HII}}{R^3}$.   For $R=70$ cMpc,  this turns out to be $\approx 11$ mK for the test bubble of size 50 cMpc and signal amplitude of 30 mK. This exercise will be important in subsequent discussions.

Now that we know that the likelihood will be maximum when the filter parameters are such that it mimics the test \HII signal, we can estimate the strength of the signal by computing the SNR for the best-fit parameters (which would be close to the true input values in this toy model). We find that the SNR of the \HII bubble which is estimated following equation (\ref{snr_bestfit}) is $\approx 200$  for 20 h of observation with SKA1-low. We also plot the SNR as a function of the radius \HII bubble for the same observation configuration as in Figure \ref{fig:snr}. We show the results for three different values of $A_{\TB}$ as shown in the figure, where smaller values of $A_{\TB}$ correspond to higher ionization fraction outside the bubble. For $A_{\TB} = 30$~mK (which corresponds to a completely neutral medium outside the bubble), the SNR for a 20h-long SKA1-low observation at redshift 7 for a bubble radius as small as $\approx 10$ cMpc will result in an SNR $\approx 10$, which indicates a highly significant detection. However, the SNR of a \HII bubble of size  $\approx$ 10 cMpc decreases to 3 for $A_{\TB}=10$ mK (which corresponds to a medium that is $\approx 67\%$ ionized outside the bubble).  The same SNR can also be achieved for a \HII bubble of radius $\approx 6$ cMpc in a completely neutral medium.

In reality, the IGM outside the bubble will clearly be partially ionized because of the presence of other galaxies, hence detection of very small bubbles could be challenging as this exercise shows. We will discuss these in more detail in Section \ref{res:real}  when we consider more realistic scenarios of reionization.

\begin{figure*}
\begin{center}
 \includegraphics[width=0.9\textwidth]{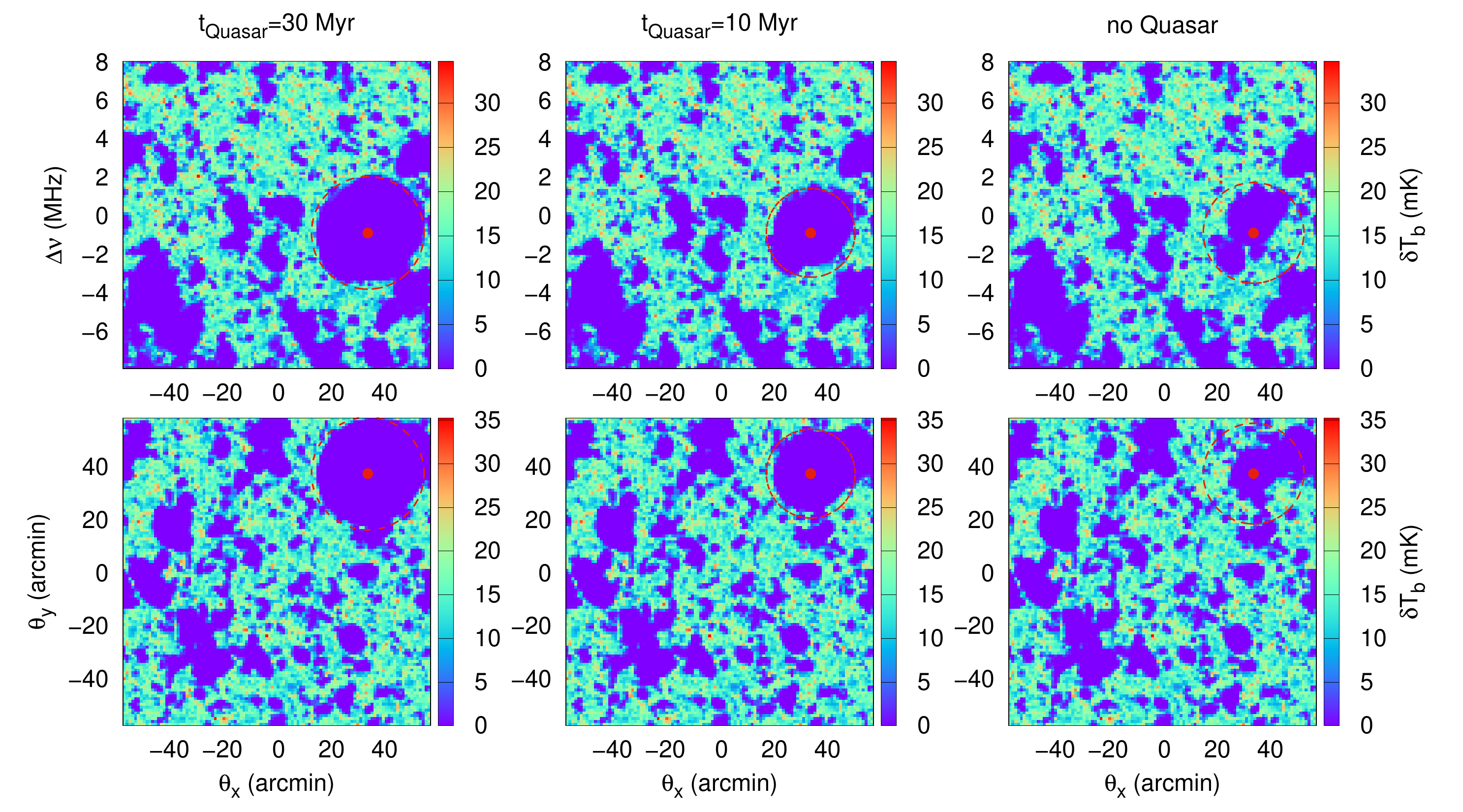}
 \caption{Two-dimensional slices of the brightness temperature obtained from the simulated light-cone. The slice is chosen so that it contains the most massive dark matter halo. The resolution of the maps is $1.075'$ which corresponds to a length scale of $\approx 2.76$ cMpc. From left to right, the columns represent models  $Galaxy-Quasar30Myr$, $Galaxy-Quasar10Myr$ and $Galaxy-noQuasar$, respectively. The bottom panels show the maps on the plane of the sky, while the top panels show the maps where the vertical axes are along the line of sight. The red dot point in each map represents the location of the rare quasar which is hosted by the most massive dark matter halo in the simulation box. The red dotted circles represent the best-fit size of the filter as estimated from the MCMC analysis.}
   \label{fig:tbmap}
\end{center}
\end{figure*}

As a final exercise with the toy model, we use the combined likelihood with MCMC-based analysis to constrain the parameters characterizing the filter (and hence the signal). In this case, we assume that we do not have any prior knowledge of the size or position of the bubble and hence vary all the five free parameters of the filter simultaneously. Such exploration of the five-dimensional parameter space of the filter not only provides the best-fit parameters but also the corresponding confidence intervals. The details of the MCMC analysis are listed in Table \ref{tab:testbub}. The best-fit parameters we obtain are $\mathbf{\hat{\mu}} =  \{50.11 \pm 0.4 ~\rm cMpc, 0.02' \pm 0.32', 0.02' \pm 0.3',  -0.002 \pm 0.03 ~\rm MHz, 29.65 \pm 1.3 ~\rm mK \}$ where the errors denote the $1\sigma$ uncertainties. These are consistent with the input parameters of the test \HII bubble whose size is 50 cMpc, coordinates of the center are ($0', 0', 0$ MHz), and amplitude of signal outside the \HII region is  30 mK. The standard deviation also gives a fair idea about the typical error-bars on the parameters for 20h of observation. The confidence intervals obtained from the MCMC analysis are shown in Figure \ref{fig:images4}. The top panel in each column shows the probability distribution function (PDF) for the corresponding parameter. One can see that the PDFs peak at the input parameter values (shown by the vertical dotted lines) of the test \HII bubble. We also note from the two-dimensional contours that there is no significant correlation between the parameters. Note that the range of parameter space plotted in this figure is smaller compared to the total prior range of parameter space explored in the analysis (see Table \ref{tab:testbub}), hence none of the constraints are affected by the priors.

\subsection{Realistic scenario}
\label{res:real}

Next, we consider several realistic scenarios of the 21~cm signal from the EoR. These scenarios are motivated by the recent detections of high-redshift quasars \citep{Mortlock11, 2015ApJ...798...28K, Venemans15,  2018Natur.553..473B}. These rare quasars are very luminous compared to the galaxies and thus, expect to create large \HII bubbles around them. In principle, one may expect that the \HII regions around those rare quasars will be distinct in size. Hence the measurement of the size of such \HII regions around the individual quasars can unravel various properties of these sources, such as the luminosity of ionizing photons produced by them, their age etc.

We model such bright quasars by assuming that there is one such source in our simulation box, and we take the most massive halo within the box at redshift 7 to be the host. Following \citet{Mortlock11}, we set the luminosity of this quasar so that it emits $1.3\times 10^{54}$ ionizing photons per second. These choices are similar as adopted in \citet{2012MNRAS.426.3178M}. In our simulation, the position of the rare quasar turns out to be  $\theta_X = 34.94'$, $\theta_Y=37.625',  \Delta\nu=-0.875$ MHz. Note that the quasar is included in addition to all other sources (stars and mini-quasars) as described in Section \ref{sec:source}. For our chosen reionization history, the ionizing photons from the galaxies set the volume averaged ionization fraction at redshift 7 to $\approx$0.4.

\begin{figure*}
\centering
\includegraphics[scale=0.35,angle=0]{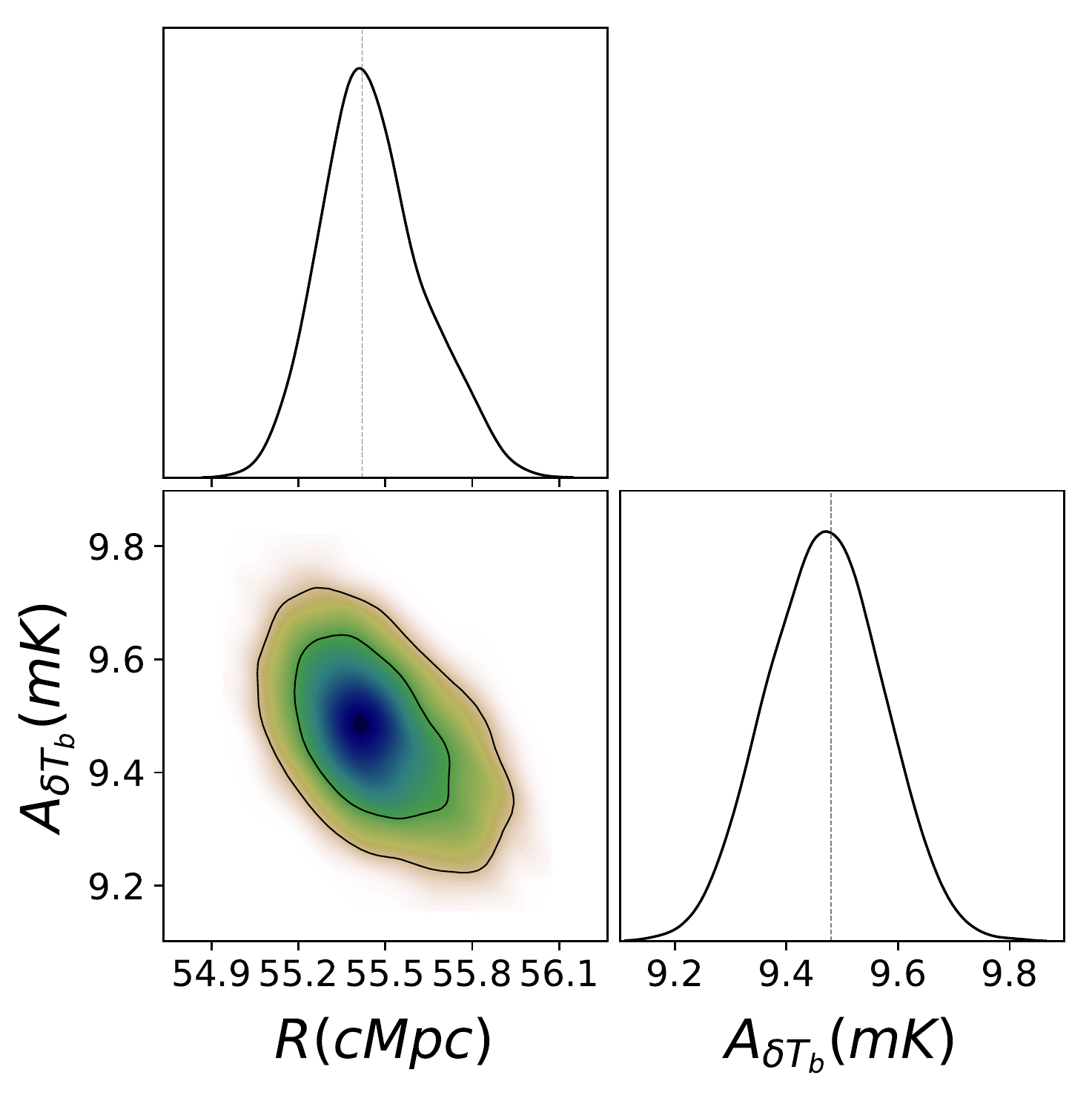}
\includegraphics[scale=0.35,angle=0]{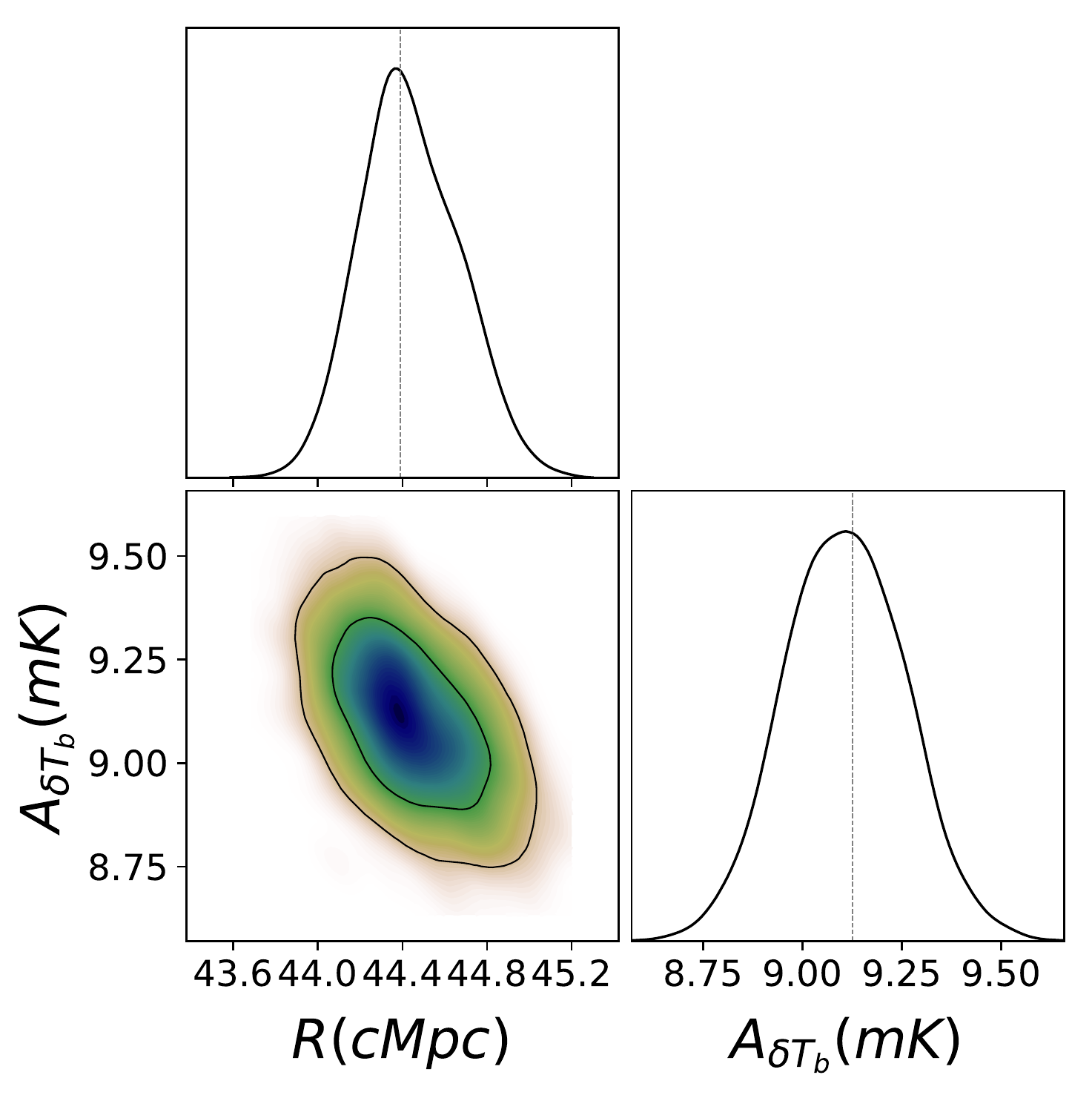}
\includegraphics[scale=0.35,angle=0]{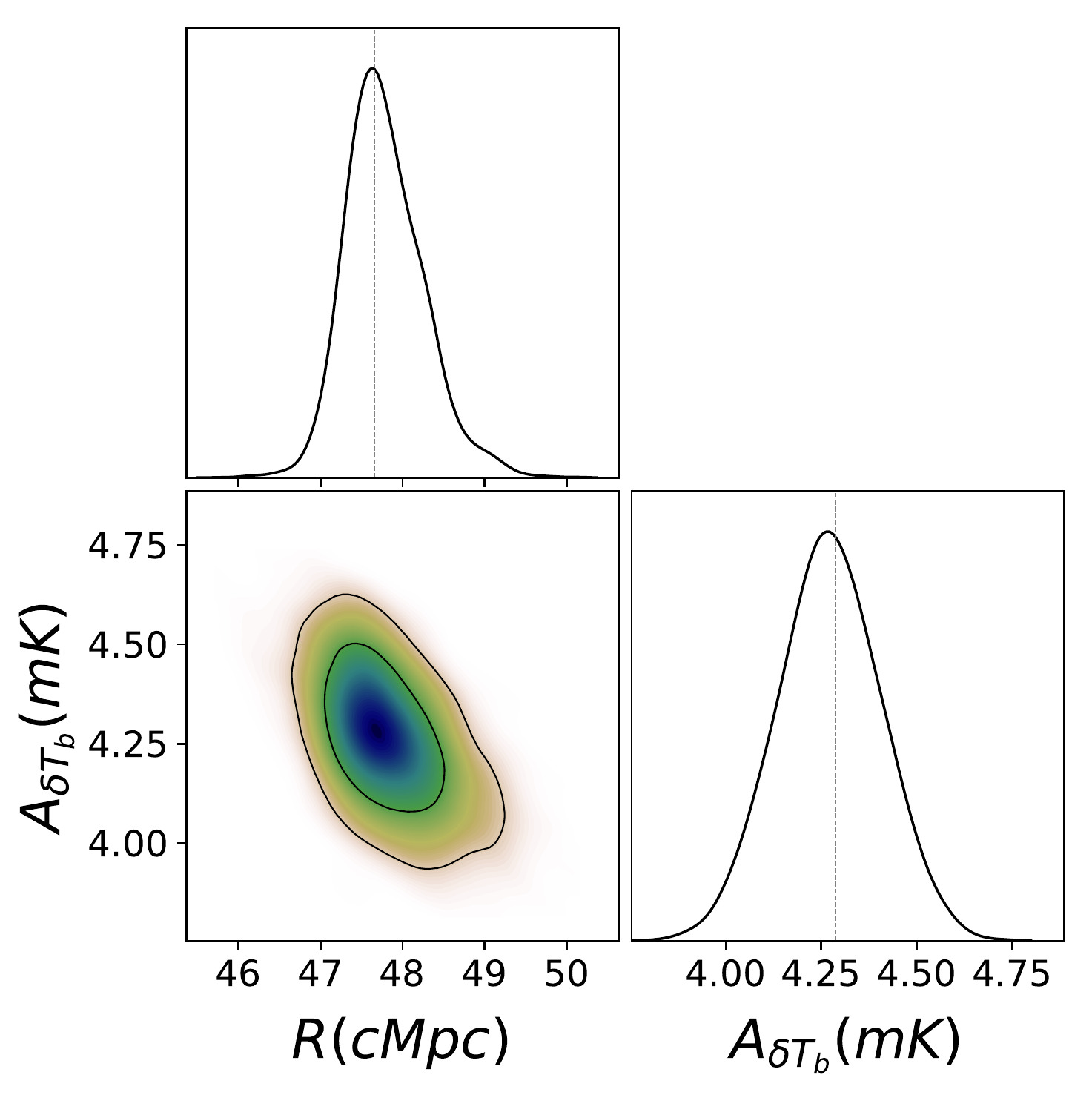}
\caption{Posterior constraints obtained from the MCMC analysis over the two-dimensional parameter space characterizing the model bubble signal. The analysis assumes that the location of the bubble is known beforehand. The contour levels in the two-dimensional contour plots represent 1$\sigma$ and 2$\sigma$ confidence levels respectively. The vertical dotted lines represent the best-fit values of the parameters obtained from the MCMC analysis. From left to right, the mean brightness temperature of the simulation boxes are 9.9, 10 and 10.1 mK respectively. The diagonal panels represent the marginalized probability distribution of each parameter. The panels from left to right represent the $Galaxy-Quasar30Myr$, $Galaxy-Quasar10Myr$ and $Galaxy-noQuasar$ models respectively. The mock observation setup is described in Table \ref{table_obs}.}
\label{fig:test_quasar}
\end{figure*}

\begin{figure*}
\centering
\includegraphics[scale=0.7,angle=0]{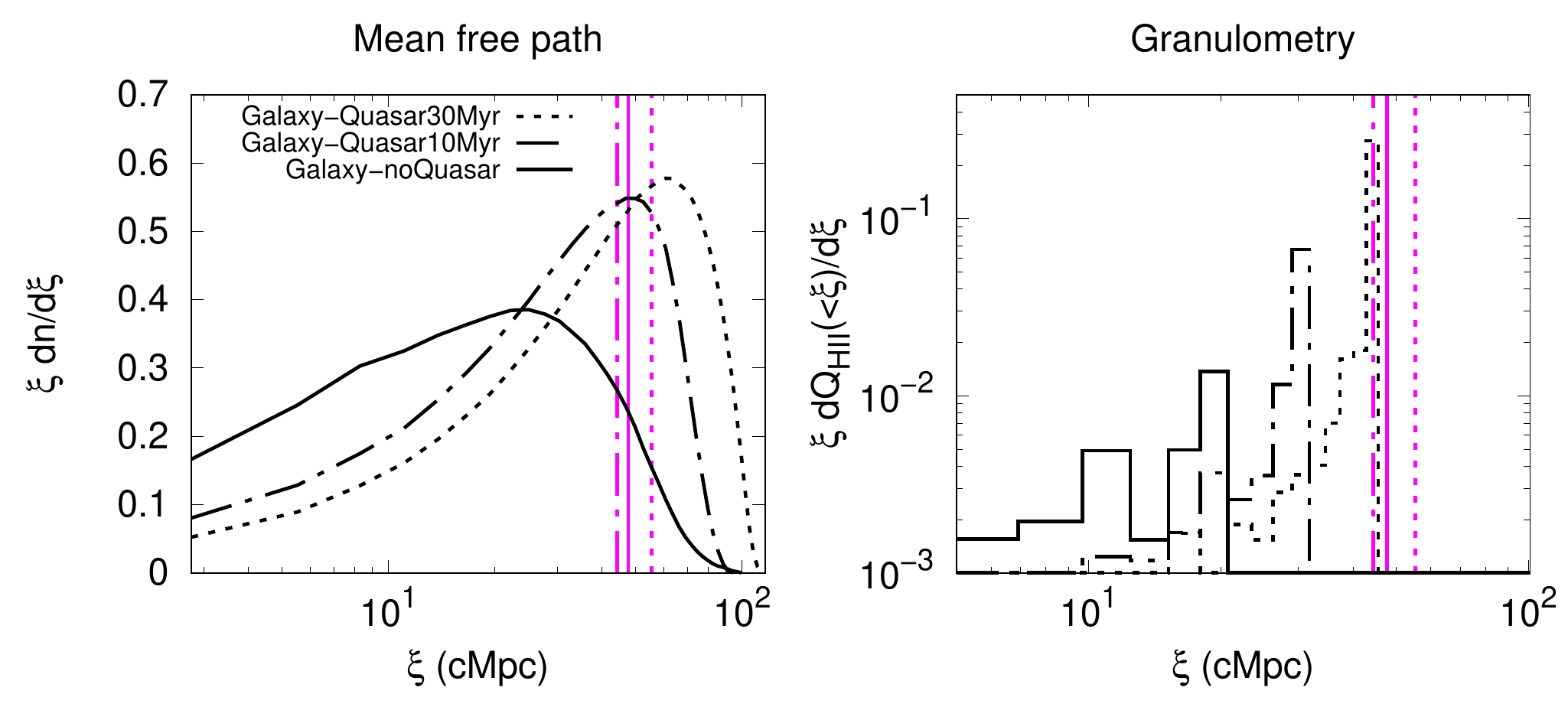}
\caption{The probability density function of the ionized region of radius $\xi$  around the most massive dark matter halo in our simulation box. Three different curves correspond to three different scenarios considered in this work. We use the mean free path and Granulometry method in the left and right panel respectively. Instead of considering the entire part of our mock sky, we consider only a spherical region around the most massive dark matter halo to estimate these bubble size distributions. The sizes of the considered regions are the same as the best-fit values of $R$ obtained from the MCMC analysis. The vertical magenta lines denote the best-fit $R$ values from the MCMC chains where we vary only $R$ and $A_{\TB}$ parameters.  }
\label{fig:bsdquasar}
\end{figure*}

\begin{table*}
\centering
\small
\tabcolsep 6pt
\renewcommand\arraystretch{1.5}
   \begin{tabular}{c c c c c}
\hline
Parameters  & Explored range & \specialcell{$Galaxy-Quasar30Myr$\\ Best fit ($\sigma$)}  & \specialcell{$Galaxy-Quasar10Myr$\\ Best fit ($\sigma$)} 	& \specialcell{$Galaxy-noQuasar$\\ Best fit ($\sigma$)}  \\
\hline
$R$ (cMpc)    & 10.0, 100.0   &  55.4 (0.18)      & 44.4 (0.23)	& 47.7 (0.48)  			\\
$A_{\TB} (\rm mK)$  & -100, 100  & 9.48 (0.1) & 9.13 (0.15) & 4.29 (0.13)\\
\hline
SNR & - & 74.6 (-) & 50.6 (-) & 26.7 (-) \\
${\TB}_{,\text{ box}} (\text{mK})$   & -  & 9.9 (-) & 10 (-) & 10.1 (-)\\
$\frac{R^{\text{ MFP}}_{\text{ peak}}}{\text{ cMpc}}$ ($\frac{R^{\text{ GN}}_{\text{ peak}}}{\text{ cMpc}}$)   & - & 60.8 (44.2) & 46.9 (30.4) & 24.8 (19.3)\\
\hline
\end{tabular}
\caption[]{Similar to Table \ref{tab:testbub}, this shows the results of the MCMC analysis for three models of the luminous quasar surrounded by realistic ionized regions. We assume that the position of the quasar is known from other experiments and vary only two parameters $R$ and $A_{\TB}$. We quote the SNR for the best-fit parameters defined in equation (\ref{snr_bestfit}).
${\TB}_{,\rm box}$ represents the average brightness temperature of the simulation box. $R^{\rm MFP}_{\rm peak}$ and $R^{\rm GN}_{\rm peak}$ are the characteristic sizes of the bubbles around the largest dark matter halo at which the PDFs of the sizes of the bubbles have peaks. The superscripts `MFP' and `GN' denote mean-free-path and Granulometry method respectively. 
}
\label{tab_mcmc2d}
\end{table*}

\begin{figure*}
\begin{center}
\includegraphics[width=0.7\textwidth]{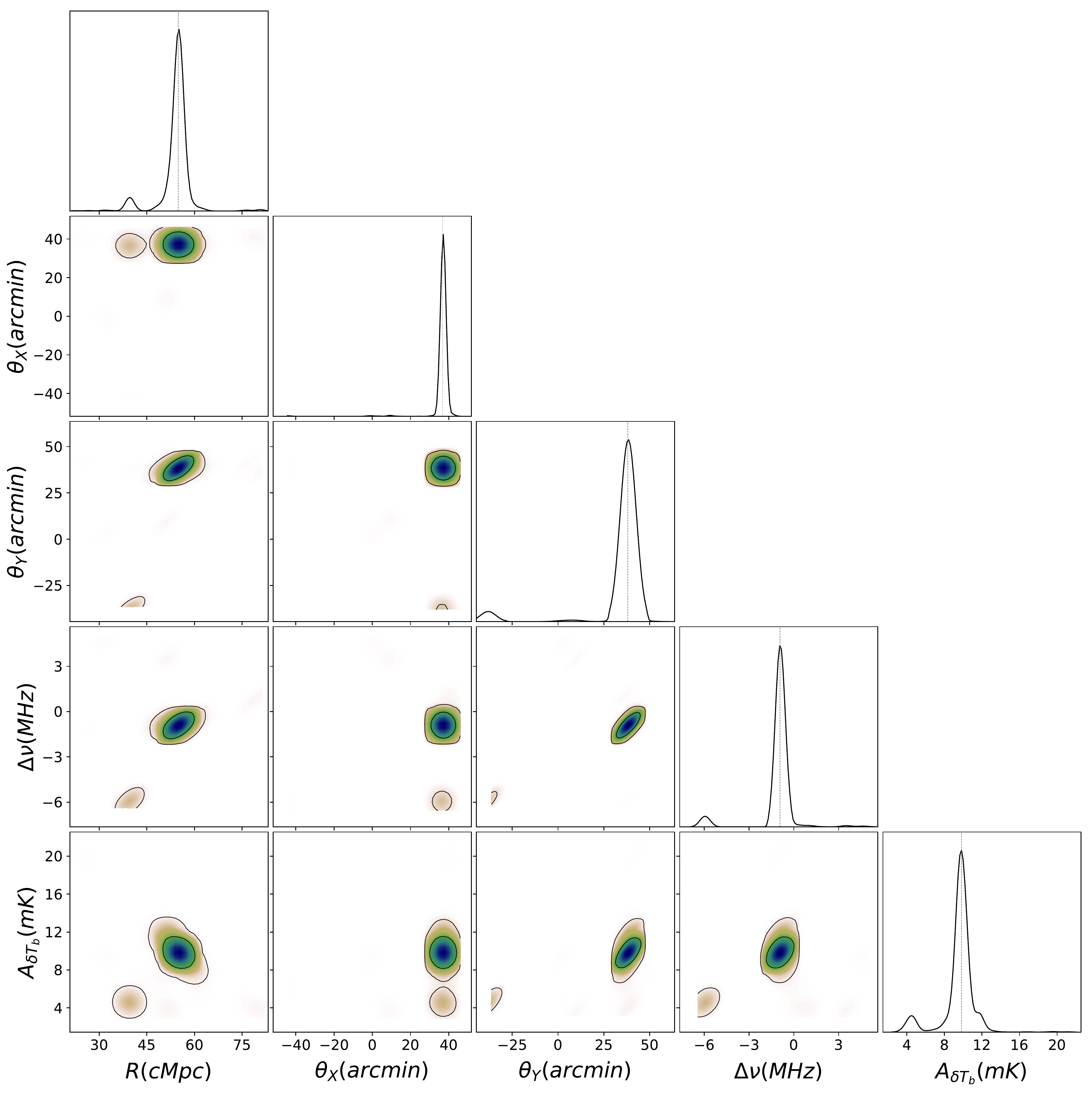}
    \caption{Posterior constraints on the realistic $Galaxy-Quasar30Myr$ model obtained using the MCMC analysis. The contour levels in the two-dimensional contour plots represent 1$\sigma$ and 2$\sigma$ confidence levels respectively. The diagonal panels represent the marginalized probability distribution of each parameter. The vertical dotted lines represent the best-fit values of the parameters obtained from the MCMC analysis. The best-fit values of the position coordinates of the filter indicate the \HII bubble around the only Quasar considered in this model whose position is $(34.94', 37.625', -0.875 ~\rm MHz)$. The mean $\TB$ of the simulation box is 9.9 mK. This analysis corresponds to 20 h of observation with SKA1-low at redshift 7.  }
    \label{fig:test_quasar30_5d}
\end{center}
\end{figure*}

\begin{figure*}
\begin{center}
\includegraphics[width=0.7\textwidth]{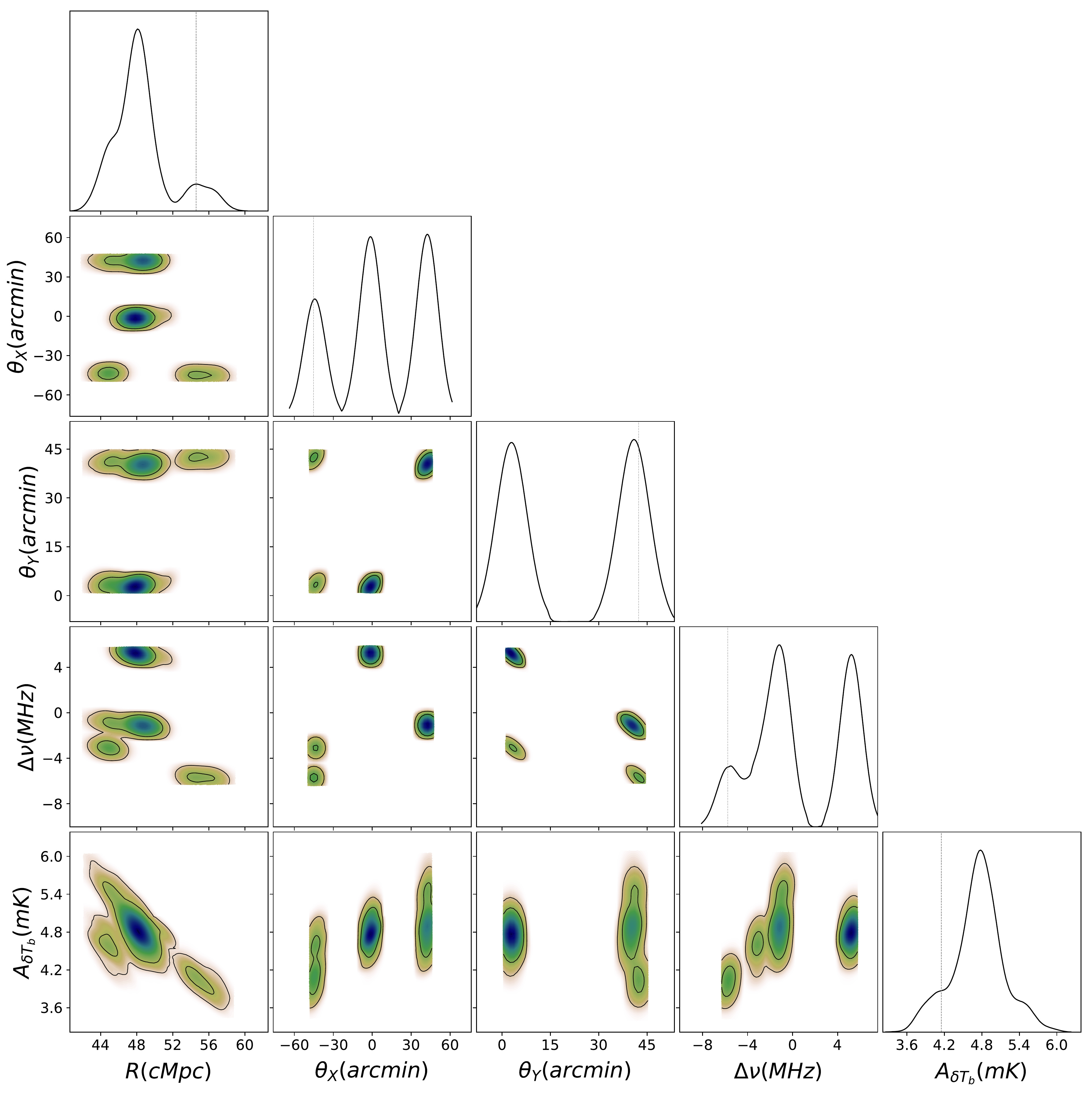}
    \caption{Posterior constraints on the realistic $Galaxy-noQuasar$ model obtained using the MCMC analysis. The contour levels in the two-dimensional contour plots represent 1$\sigma$ and 2$\sigma$ confidence levels respectively. The diagonal panels represent the marginalized probability distribution of each parameter. The vertical dotted lines represent the best-fit values of the parameters obtained from the MCMC analysis. The mean $\TB$ of the simulation box is 10.1 mK. This analysis corresponds to 20 h of observation with SKA1-low at redshift 7.}  
    \label{fig:test_noquasar_5d}
\end{center}
\end{figure*}

To test the performance of our likelihood-based framework, we choose three different models for the bright quasar as follows:
\begin{enumerate}
    \item $Galaxy-Quasar30Myr$: In this model, we assume the lifetime of the quasar is 30 Myr at redshift 7.
    \item $Galaxy-Quasar10Myr$: This is similar to the previous scenario except that the lifetime of the quasar is taken to be 10 Myr at redshift 7.
    \item $Galaxy-noQuasar$: In this model, there are \emph{no} bright quasars in the box. Hence the box only contains ionized regions generated by the stars and mini-quasars as described in section \ref{sec:source}. 
\end{enumerate}

In the bottom panels of Figure \ref{fig:tbmap}, we show the two-dimensional maps of $\TB$ on the sky plane with the slice chosen such that it contains the most massive dark matter halo in the box. The top panels of the same figure show the maps where the vertical axes is along the frequency direction. The three panels from left to right correspond to the three scenarios mentioned above. The volume average brightness temperature of these three scenarios are 9.9 mK, 10 mK, and 10.1 mK respectively, the difference being due to the size of the ionized region produced by the luminous quasar. The \HII region around the quasar is distinctly larger compared to the other \HII bubbles in the middle and left panel of the figure. Also, the size of the \HII region around the quasar increases with the lifetime. It is also interesting to note that the ionized region around the most massive halo is larger compared to the other such regions at the same redshift or at same $\Delta \nu$ even in the absence of any luminous quasar (see the right panel of Figure \ref{fig:tbmap}). These agree with previous results of \citet{2012MNRAS.424..762D, 2013MNRAS.429.1554F, 2017MNRAS.468.3718K}.  Although there are no luminous quasars in the box,  the largest \HII regions nevertheless are expected to be around the massive haloes which form earlier in time and hence contain more number of stars than the less massive haloes. Also, these massive haloes are strongly clustered, hence the overlap of the \HII bubbles is quite efficient  leading to larger \HII regions. On the other hand, note that the sizes of the \HII regions consistently increase as $\Delta \nu$ decreases. This is because a part of the map with a smaller $\Delta \nu$ corresponds to a lower redshift and thus corresponds to a later stage of EoR where the characteristic sizes of the \HII regions increase. This is also known as the light-cone effect \citep{ghara15b}.

Next, we apply our parameter estimation method to the visibilities of these maps. The specifications of the observations are the same as previously used in the toy model of the isolated \HII region in the previous section (see also Table \ref{table_obs}). We first take the case where the location of the quasar is known from, say, optical or IR observations. In this case, three of the five parameters $\theta_X, \theta_Y$ and $\Delta \nu$ are fixed beforehand  and we perform the MCMC analysis by varying only two parameters $R$ and $A_{\TB}$. We fix the coordinates $\theta_X=34.94'$, $\theta_Y=37.625'$ and $\Delta \nu=-0.875$ MHz. Note that under this approach, the model $Galaxy-noQuasar$ is only of academic interest as there is no quasar at the target location.

The results of the two-parameter MCMC analysis for these three scenarios are shown in Figure \ref{fig:test_quasar} and listed in Table \ref{tab_mcmc2d}. The best-fit value of $R$ as obtained from the MCMC analysis for the $Galaxy-Quasar30Myr$ model is $R\approx 55$ cMpc which is visually similar as shown in the left panel of Figure \ref{fig:tbmap}. The best-fit regions are highlighted by the dashed circles in the figure. We also find that the best fit value (standard deviation) of $A_{\TB}$ is $\approx 9.5 ~(0.1)$ mK which is smaller than the mean of the brightness temperature of the box (which is 9.9 mK in this case). In fact, the mean of the brightness temperature of the box lies well outside of the credible limits (more than $3-\sigma$ away) of $A_{\TB}$.  Hence our model is unable to recover the input true value of $A_{\TB}$.
The reason for this bias in the recovered amplitude is that the spherical filter is unable to describe the complex shape of the ionized bubble and includes some neutral regions. We also note from the table that the precision of these recovered parameters is $\lesssim 1\%$.

The conclusions are similar for the other model including the quasar, i.e., $Galaxy-Quasar10Myr$. The best-fit values (standard deviations) of $R$ and $A_{\TB}$ are 45 (0.23) cMpc and 9.1 (0.15) mK respectively for this case. In this case too, the mean $\TB$ of the simulation box (which is $\approx$ 10 mK) is well outside the credible limits of our recovered $A_{\TB}$.

Unlike these two scenarios, the best-fit parameters for the scenario $Galaxy-noQuasar$ is not straightforward to interpret. We find that the best-fit parameter values (standard deviations) of $R$ and $A_{\TB}$ are $\approx 48$ (0.5) cMpc and $\approx 4.3$ (0.13) mK, respectively in this case. The best-fit size of the filter appears to be larger than the expected size of the \HII region around the most massive halo (see the right panels in Figure \ref{fig:tbmap}). On the other hand, the best fit value of $A_{\TB}$ is smaller than ${\TB}_{,\rm box}$ (and also the best fit value of $A_{\TB}$ as obtained from other two scenarios) by a factor of 2.5. The main reason why the recovery is significantly worse in this case is that the \HII regions around the chosen point are highly non-spherical in shape, while the filter we are using to model the region is perfectly spherical. As a result, there are no set of parameters for which the filter overlaps significantly with the ionized region (the overlap is somewhat better in the other two models). The maximum likelihood corresponds to a filter of size $R\approx 48$ cMpc in this case. A region of this size around the prefixed position in the mock map contains a mixture of ionized and neutral regions. Remember that we have encountered a similar situation in section \ref{res:iso_HII} also. Bringing the same analogy, we can represent the best-fit value for  $A_{\TB} \simeq ({\TB}_{,\rm box}\times \frac{R^3_{\rm eff,HII}}{R^3}) \simeq 4.3$ mK where ${\TB}_{,\rm box} \approx 10.1 ~\rm mK$, $R\approx 48 ~\rm cMpc$. This means the effective size of the \HII regions within the spherical regions of radius $\approx 48$ cMpc around the most massive halo is $R_{\rm eff,HII}\approx 35$ cMpc. The effective sizes of the \HII regions for the best-fit values of $Galaxy-Quasar30Myr$ and $Galaxy-Quasar10Myr$ models are $\approx$ 53 cMpc and 41 cMpc respectively.

Interestingly, the best-fit $R$ obtained for the $Galaxy-noQuasar$ is very similar to the other two models. This indicates that the size the ionized region, as detected by the spherical filter, cannot distinguish between scenarios where the quasar is present against those where it is absent.

We will now compare these best-fit values of $R$ with the size estimated using methods such as mean-free-path (MFP \citep{mesinger07} and Granulometry method (GN) \citep{2017MNRAS.468.3718K}. For this, we choose a part of the simulation box around the most massive dark matter halo within a radius given by the best-fit values of these scenarios. Note that many of the large \HII bubbles at this stage of reionization are inter-connected, hence it is important to restrict the analysis to a limited region of interest. We consider all the ionized (with $\XHII>0.5$) grid points within this selected part of the box. The bubble size distributions for these regions are shown in Figure \ref{fig:bsdquasar}. The sizes $R^{\text{ MFP}}_{\text{ peak}}$ and  $R^{\text{ GN}}_{\text{ peak}}$ at which the probability distribution functions become maximum for both the methods are distinctly smaller than the best-fit $R$ value for the $Galaxy-noQuasar$ case (also see Table \ref{tab_mcmc2d}). On the other hand, best-fit $R$ values we obtain are well bracketed by the $R^{\text{ MFP}}_{\text{ peak}}$ and  $R^{\text{ GN}}_{\text{ peak}}$ values for  the $Galaxy-Quasar30Myr$ and $Galaxy-Quasar10Myr$ cases. The difference between the characteristic size estimated from these two methods and the best-fit $R$ value is the smallest for the case of $Galaxy-Quasar30Myr$ as the \HII bubble is more spherical in this case. For an irregular shaped \HII region, the best-fit $R$ value obtained from our framework will be larger than the size estimated using the MFP and GN method. This is because our filter-based method takes into account the volume in all the \HII regions within $R$, while the MFP and GN methods follow individual smaller \HII regions and the details of their structure. To check this further, we estimate the effective radius of the ionized region within the chosen part of the box, i.e., we compute $R_{\rm eff} = (\frac{3\times V_{\rm ion}}{4\pi)})^{1/3}$ where $V_{\rm ion}$ is the total volume of the ionized regions in the selected part of the box. We found $R_{\rm eff}$ values are $\approx$ 50, 39 and 35 cMpc for the  $Galaxy-Quasar30Myr$, $Galaxy-Quasar10Myr$ and $Galaxy-noQuasar$ cases respectively. These are close to the  $R_{\rm eff, HII}$ values $\approx$ 53, 41, and 35 cMpc estimated previously using the best-fit parameter values.

The SNR corresponding to the best-fit parameters for the  $Galaxy-Quasar30Myr$, $Galaxy-Quasar10Myr$ and $Galaxy-noQuasar$ models are 74.6, 50.6 and 26.7 respectively. Note that the SNR is lower in case of $Galaxy-noQuasar$ although the best fit size of the \HII region is larger than the $Galaxy-Quasar10Myr$ model. This is due to the smaller best fit value of the parameter $A_{\TB}$ in the $Galaxy-noQuasar$ model. In fact, a smaller SNR in the $Galaxy-noQuasar$ model is expected as the size of the \HII bubble is smaller which can be seen in the maps. Also, note that all these SNRs are smaller compared to the SNR estimated for the toy \HII bubble model. This is due to a smaller $A_{\TB}$ value in these realistic scenarios where the region outside the bubbles are significantly ionized by other sources.

From the above discussion, it is clear that the recovered best-fit value of $R$ cannot distinguish between scenarios where the quasar is present against those where it is absent. However, we found that the SNR is considerably higher in the presence of a quasar compared to the one without a quasar. In principle, one can use a threshold on the SNR to distinguish between these two scenarios, although it is not straightforward to estimate that threshold. This needs further investigation which is out of the scope of this paper and will be addressed in the future.

Next, we will consider the full five-dimensional parameter space study for the scenario $Galaxy-Quasar30Myr$. Figure \ref{fig:test_quasar30_5d} shows the parameter constraints from this MCMC analysis. The best-fit parameter values (standard deviations) obtained from the analysis are $R=55.22 (5.04) ~\text{ cMpc}$, $\theta_X = 36.9' (5.46')$, $\theta_Y= 38.2' (17.1')$, $\Delta\nu=-0.95 (1.28) ~\text{ MHz}$ and $A_{\TB}=9.71  (1.7) ~\text{ mK}$. Clearly, these best-fit values of the parameters correspond to the largest \HII bubble in the field of view which is the ionized region around the quasar as shown in the left panel of Figure \ref{fig:tbmap}. For reference, note that the location of the most massive halo corresponds to $\theta_X = 34.94'$, $\theta_Y = 37.625'$ and $\Delta \nu = $-0.875 MHz. Thus the recovered location of the bubble centre does not exactly correspond to the halo location, however, the deviation is $\lesssim 5\%$. The SNR obtained from these best-fit values of the filter parameters is $\approx$ 76 which is very similar to what we found in the case where the location of the quasar was assumed to be known. Re-doing the same analysis for even smaller observation time $t_{\rm obs}\approx 5$ h, we find the best fit values approach the expected values quite well with a SNR $\approx 36$.

When we vary all the parameters of the $Galaxy-Quasar10Myr$ model, the best-fit parameter values (standard deviations) obtained are $R=44.57 (4.9) ~\text{ cMpc}$, $\theta_X = 38.2' (23.')$, $\theta_Y= 39.3' (21.5')$, $ \Delta\nu=-0.858 (2.9) ~\text{ MHz}$ and $A_{\TB}=9.17  (2.) ~\text{ mK}$ which correspond to a SNR of $\approx$ 51. These are similar to what we found when we varied only two parameters $R$ and  $A_{\TB}$ assuming that the source location is known. On the other hand, the best-fit parameters for the $Galaxy-noQuasar$ model are  $R=54.6 (3.1) ~\rm cMpc, \theta_X = -45.2' (33.8'), \theta_Y= 42.4' (19'), \Delta\nu=-5.76 (3.9) ~\rm MHz$ and $A_{\TB}=4.15  (0.4) ~\rm mK$. 

We present the results from the MCMC analysis for this scenario in Figure \ref{fig:test_noquasar_5d}. The SNR for this best-fit values is $\approx$ 32 which is slightly larger than the SNR of the bubble around the most massive halo, i.e., 26.7. In this case, the inferred location is different than the location of the most massive dark matter halo. In fact, the best-fit value  $ \Delta\nu=-5.76  ~\rm MHz$ means the location of the best-fit bubble corresponds to a redshift which is smaller than 7. Due to the light-cone effect, this \HII region is larger than the one around the most massive halo at redshift 7. However, the posterior distribution as shown in Figure \ref{fig:test_noquasar_5d} shows multiple convergence points which means the observed part of the sky contains multiple \HII regions with similar significance. After pointing out the most probable positions of the large \HII bubbles from the posterior distribution, one can perform another MCMC analysis within a limited part of the sky around those points. In this way, a more accurate characterisation of the \HII regions could be possible.  The multimodal posterior distribution for $Galaxy-noQuasar$ model is distinctly different than other two models that contain Quasars. Thus, in principle, the absence of a strong multimodal posterior could be used to distinguish a field with a quasar.

To summarize the results, we find that our Bayesian likelihood-based framework should be able to measure the locations and sizes of the largest \HII bubbles and also an estimate of the neutral fraction outside at redshift $\approx 7$ even for a shallow observation with the SKA1-low. Of course, this framework has a limitation because the \HII region we are trying to detect is modelled as spherical while the actual shapes could be quite different. In this sense, our method cannot provide all the details of the signal which, e.g., tomographic maps can provide. However, it is important to keep in mind that the detection of the \HII region directly in images requires a much longer observation time. For example, a $10 \sigma$ detection of the peak signal through images with resolution $\approx 10'$ and bandwidth 19 MHz at redshift 7 requires at least $\approx$ 60 h of SKA1-low observation time \citep{ghara16}. On the contrary, our filter-based method is able to detect the \HII bubbles using much less observation time, say, between 5-20 h. This thus motivates a new observation strategy which consists of two steps: The first step is to detect a large \HII bubble by applying the filters on the observed visibilities in a shallow observation with a few hours of observation time with SKA1-low. As these large \HII regions may be rare, several such shallow observations should be done towards the different field of views. This strategy will also be useful to avoid field with complex features from the foregrounds, ionosphere etc. 
Our analysis also suggests that regions with SNR $\gtrsim 50$ are likely to be ionized by a quasar while regions with smaller SNR do not contain any active quasar.
Once, a tentative detection of the signal in some field is done with a significant SNR, the next step will be to observe that field for a longer time. These deeper observations will be useful to visually distinguish the \HII regions in images and characterize the signal with more detail.

\section{Conclusions \& Discussions}
\label{sec:con}
In this study, we introduce a Bayesian approach to explore the possibility of constraining the properties of individual ionized regions during the Epoch of Reionization using the upcoming SKA1-low. Such individual large ionized regions are expected to contain clusters of UV emitting sources such as early galaxies or even rare early Quasars. Thus such detection of ionized regions using 21~cm observations can trigger followup experiments in Infrared to characterise the sources at the center of such \HII bubbles. Our framework closely follows the calculations of \citet{1992PhRvD..46.5236F} meant for detecting and measuring parameters of gravitational wave. We extend the matched filtering technique of \citet{kanan2007MNRAS.382..809D,2012MNRAS.426.3178M} to find a more rigorous method to obtain the posterior distribution of the parameters that characterize the ionized bubble given an input observational data.

In this work, we assume that the signal from the bubble can characterized by a spherical top-hat filter having five parameters, namely, the radius and the three position coordinates of the centre of the sphere and an amplitude that measures the contrast between the signal inside and outside the \HII region. This spherical signal is used to define the likelihood which is then coupled to an MCMC-based analysis to extract the parameters of the \HII regions from the simulated mock observations appropriate for the SKA1-low. This entire framework is suitable for both targetted and also blind searches of the \HII regions during the EoR. The main findings of this study are listed below.

\begin{itemize}

\item When we test the framework on a toy model involving a single isolated \HII bubble in an uniform medium, the method accurately estimates the size and position of the bubble. In addition, it also determines the average brightness temperature of the signal accurately. The SNR, which measures the strength of the detection, increases with the size of the \HII bubble. The SNR of the test \HII bubble of size $\approx 50$ cMpc at redshift $\approx 7$ is $\approx 200$ for 20 hours of observation time with SKA1-low when the medium outside is completely neutral. For the same observation, the SNR $\approx 10$ for a \HII bubble of size as small as $\approx 10$ cMpc. 

\item When we apply the same framework to more realistic reionization maps, the framework determines the position and size of the largest \HII bubble in the FOV reasonably accurately.

\item The bubble properties can be recovered more accurately when the bubble is large and close to spherical in shape. This is the case, e.g., when the quasar has a relatively long lifetime.  In the case of an irregular shaped \HII region, the estimated characteristic size of the bubble can be larger than the real size. In that case, the contrast between the averaged brightness temperatures of the entire volume and regions inside the bubble turns out to be smaller. 

\item We find the SNR of the \HII region around a typical quasar as observed in \citet{Mortlock11} is $\approx 70$ for 20 hours of observation with SKA1-low. However, the SNR varies with the size of the bubble which, in turn, depends on the properties of the quasar and the surrounding IGM. Thus, we argue that the estimated size of the \HII region around a quasar using our method can provide information about the properties of the quasar. The value of SNR can also be useful in distinguishing between cases where the regions are ionized by quasars against those which are ionized by stars in galaxies.

\end{itemize}

While we find that the detectability of the large \HII regions during the EoR is significant when we apply this method to the observed visibilities, we should also keep in mind the various assumptions that go into this study. This study assumes that the observed visibilities are accurately calibrated and all artifacts have been removed. In reality, such a favorable scenario is difficult to achieve. In addition, we assumed that the Galactic and extra-galactic foregrounds are well behaved and can be removed accurately from the visibilities. In principle, our method can be extended such that it can be applicable to visibilities with the well-behaved foregrounds. This may reduce the SNR slightly. We will address the performance of the pipeline in the presence of foregrounds in future studies.

While our present work is focussed on detecting the large \HII regions during the epoch of reionization, the same formalism is also applicable to find emission/absorption regions during the Cosmic Dawn. During the Cosmic Dawn, the IGM is expected to consist of large emission/absorption regions around the sources \citep[see e.g.,][]{ghara15a, ghara16, 2019MNRAS.tmp.1183R}. Such blind searches can indicate the possible positions of sources at such high redshifts and thus, will be relevant for future NIR observations with instruments such as the JWST.

\section*{Acknowledgements}
We would like to thank an anonymous referee for insightful comments that helped us to improve the paper. The authors would like to thank Kanan K Data, Samir Choudhuri, Garrelt Mellema, Sambit Giri for useful discussions and comments on this work. TRC acknowledges support from the Associateship Scheme of ICTP, Trieste.

\bibliography{matched_filter}


\appendix

\section{Detailed derivation of the likelihood}
\label{appen:like}

We provide details of the derivation of the form of the likelihood used in this work to constrain the parameters. Our method closely follows that of \citet{1992PhRvD..46.5236F} where the calculations were meant for detection and interpretation of the gravitational wave signal. We repeat the formalism here for the radio visibilities.

Consider a visibility measurement $V(\vec{U}, \nu)$ using a radio interferometer with baselines $\vec{U}$ and operating frequency $\nu$. The signal is assumed to have two components, namely, the signal $S(\vec{U}, \nu)$ and the system noise $N(\vec{U}, \nu)$. If the noise component dominates the measurement, it becomes difficult to detect the signal directly. In that case, we assume a model $S_f(\vec{U}, \nu; \mathbf{\mu})$ for the signal which depends on several unknown model parameters denoted by the multi-dimensional vector $\mathbf{\mu}$. We can then write the total visibility as
\be
V(\vec{U}, \nu) = \left\{
\begin{array}{ll}
S_f(\vec{U}, \nu; \mathbf{\mu}) + N(\vec{U}, \nu) & \mbox{if signal  present}, \\
N(\vec{U}, \nu) & \mbox{if signal not present}.
\end{array}
\right.
\ee

Now, while calculating the detectability of the signal, we are usually interested in the quantity $P(S_f | V)$ defined as the conditional probability that the signal $S_f(\vec{U}, \nu; \mathbf{\mu})$, for unknown $\mathbf{\mu}$, is present given the observed $V(\vec{U}, \nu)$. If this probability $P(S_f | V)$ exceeds a pre-decided detection threshold, then we assume to have detected the signal. Using Bayes' theorem, we can write
\be
P(S_f | V) = \f{P(V | S_f)~P(S_f)}{P(V)},
\ee
where $P(V | S_f)$ is the probability of measuring $V$ if $S_f$ is present, $P(S_f)$ is the prior that the signal $S_f$ is present and $P(V)$ is the probability that $V$ is observed. One can express $P(V)$ in terms of two probabilities: $S_f$ present and $S_f$ absent:
\be
P(V) = P(V | 0) P(0) + P(V | S_f) P(S_f),
\ee
where $P(0)$ is the prior that the signal is \emph{not} present and $P(V | 0)$ is the probability of observing $V$ in absence of the signal\footnote{Note that in conventional likelihood calculations, the quantity $P(V)$ corresponds to the probability of observing the data which is expected to be independent of the model. Hence this quantity is fixed by normalization condition. In our case, $P(V)$ is implicitly dependent on the signal we are trying to detect.}. Further, we can write the probability $P(V | S_f)$ that the signal is present as an integral over probabilities that the signal is characterized by a particular parameter set $\mathbf{\mu}$:
\be
P(V | S_f) =  \int \de\mathbf{\mu}~p(\mathbf{\mu})~P[V | S_f(\mathbf{\mu})].
\ee
In the expression above, $P[V | S_f(\mathbf{\mu})]$ is the probability density of observing $V$ assuming $S_f(\mathbf{\mu})$ with a particular $\mathbf{\mu}$ is present and $p(\mathbf{\mu})$ is the prior probability density that $S_f$ is characterized by $\mathbf{\mu}$. 

The above equations can be manipulated to obtain
\be
P(S_f | V) = \f{\Lambda}{\Lambda + P(0) / P(S_f)},
\ee
where
\be
\Lambda \equiv \f{P(V | S_f)}{P(V | 0)} = \int \de\mathbf{\mu}~\Lambda(\mathbf{\mu}),
\ee
and
\be
\Lambda(\mathbf{\mu}) \equiv p(\mathbf{\mu}) \f{P[V | S_f(\mathbf{\mu})]}{P(V | 0)}.
\label{eq:lambda_mu}
\ee

If we are interested in measuring the values of the parameters $\mathbf{\mu}$, then we need to compute the the conditional probability $p[S_f(\mathbf{\mu}) | V]$ that the particular signal $S_f(\vec{U}, \nu; \mathbf{\mu})$ characterized by $\mathbf{\mu}$ is present in the data $V(\vec{U}, \nu)$. Note that this probability is given by
\be
p[S_f(\mathbf{\mu}) | V] = \f{\Lambda(\mathbf{\mu})}{\Lambda + P(0) / P(S_f)}
\ee
such that the integral $\int \de\mathbf{\mu}~p[S_f(\mathbf{\mu}) | V] = P(S_f | V)$. Since $p[S_f(\mathbf{\mu}) | V] \propto \Lambda(\mathbf{\mu})$ and the denominator is independent of $\mathbf{\mu}$, the probability distribution is essentially determined by $\Lambda(\mathbf{\mu})$. For example, to obtain the best-fit parameters $\mathbf{\hat{\mu}}$, it is sufficient to maximize the quantity $\Lambda(\mathbf{\mu})$ (or equivalently, $\ln \Lambda(\mathbf{\mu})$).

\subsection{Explicit form of $\Lambda(\mathbf{\mu})$}

Let us now work out the explicit form of the conditional probability defined in \eqn{eq:lambda_mu} in terms of the telescope properties. To do this, first note that the conditional probability of measuring $V(\vec{U}, \nu)$ when the particular signal $S_f(\vec{U}, \nu; \mathbf{\mu})$ is present is the same as the conditional probability of measuring $V'(\vec{U}, \nu) = V(\vec{U}, \nu) - S_f(\vec{U}, \nu; \mathbf{\mu})$ when the signal $S_f(\vec{U}, \nu; \mathbf{\mu})$ is \emph{not} present in $V'(\vec{U}, \nu)$, i.e.,
\be
P[V | S_f(\mathbf{\mu})] = P[V - S_f(\mathbf{\mu}) | 0].
\ee
Hence it is sufficient for us to work out the form of $P(V | 0)$.

In absence of the signal, the visibility is simply given by the noise $V(\vec{U}, \nu) = N(\vec{U}, \nu)$ which, in our case, is assumed to be a realization of a gaussian random field. Let us assume that there are $N_B$ baselines labelled by $\vec{U}_i,~i = 1,2,\ldots,N_B$, and $N_C$ frequency channels $\nu_{\alpha},~\alpha = 1,2,\ldots,N_C$. The visibilities can then be labelled as $V_{i \alpha} \equiv V(\vec{U}_i, \nu_{\alpha})$. We can then write the probability as
\be
P(V_{i \alpha} | 0) = P(N_{i \alpha}) = \f{1}{\sqrt{2}~ \pi \sigma_{N, i \alpha}} \exp\left(-\f{|V_{i \alpha}|^2}{\sigma_{N, i \alpha}^2}\right),
\ee
where $\sigma_{N, i \alpha}$ is the rms of the noise in the baseline $i$ and frequency channel $\alpha$.

Assuming the visibilities $V_{i \alpha}$ for different $i, \alpha$ to be independent, we can write
\be
P(V | 0) = \f{1}{\sqrt{2}~ \pi \prod_{i, \alpha} \sigma_{N, i \alpha}} \exp\left(-\sum_{i, \alpha} \f{|V_{i \alpha}|^2}{\sigma_{N, i \alpha}^2}\right).
\ee

At this point, let us shift to the continuum limit with the identification
\begin{align}
  \sum_{i, \alpha} \left(\ldots\right) &\longrightarrow \int \de^2 U \int \de \nu~n_B(\vec{U}, \nu)~\left(\ldots\right)
                                         \nline
                                         &=  N_B N_C\int \de^2 U \int \de \nu~\rho_B(\vec{U}, \nu)~\left(\ldots\right),
\end{align}
where $n_B(\vec{U}, \nu)~\de^2 U~\de \nu$ gives the number of baselines in the interval $\de^2 U~\de \nu$ and $\rho_B(\vec{U}, \nu)$ is the corresponding normalized distribution
\be
\int \de^2 U \int \de \nu~\rho_B(\vec{U}, \nu) = 1.
\ee
Hence the ratio that appears in the expression for $\Lambda(\mathbf{\mu})$ is given by
\begin{align}
  \f{P[V | S_f(\mathbf{\mu})]}{P(V | 0)} &= \f{P[V - S_f(\mathbf{\mu}) | 0]}{P(V | 0)}
                                           \nline
  &= \exp\left[N_B N_C \int \de^2 U \int \de \nu~\f{\rho_B(\vec{U}, \nu)}{\sigma_N^2(\vec{U}, \nu)} \right.
    \nline
  &\times \left\{ V(\vec{U}, \nu) S_f^*(\vec{U}, \nu; \mathbf{\mu}) + V^*(\vec{U}, \nu) S_f(\vec{U}, \nu; \mathbf{\mu}) \right.
    \nline
    & \left. \left. - |S_f(\vec{U}, \nu; \mathbf{\mu})|^2 \right\} \right].
\end{align}

Now, since the quantities in the image plane are real, the visibilities must satisfy the following constraints
\be
V^*(\vec{U}) = V(-\vec{U}),~S_f^*(\vec{U}) = S_f(-\vec{U}).
\ee
Also, since the baselines are determined by positions of antenna pairs, the corresponding distribution must be symmetric $\rho_B(\vec{U}) = \rho_B(-\vec{U})$. The same argument works for the noise rms $\sigma_N^2(\vec{U}) = \sigma_N^2(-\vec{U})$. These conditions simplify the expression as
\begin{align}
  \Lambda(\mathbf{\mu}) &= p(\mathbf{\mu})~\exp\left[N_B N_C \int \de^2 U \int \de \nu~\f{\rho_B(\vec{U}, \nu)}{\sigma_N^2(\vec{U}, \nu)} \right.
                          \nline
                          &\times \left.\left\{2 V(\vec{U}, \nu) S_f^*(\vec{U}, \nu; \mathbf{\mu}) - |S_f(\vec{U}, \nu; \mathbf{\mu})|^2 \right\} \right].
\end{align}
The above expression can be further simplified if the rms noise is independent of the baseline and frequency channel:
\begin{align}
  \Lambda(\mathbf{\mu}) &= p(\mathbf{\mu})~\exp\left[\f{1}{\sigma_{\rm rms}^2} \int \de^2 U \int \de \nu~\rho_B(\vec{U}, \nu) \right.
                          \nline
                          &\times \left.\left\{2 V(\vec{U}, \nu) S_f^*(\vec{U}, \nu; \mathbf{\mu}) - |S_f(\vec{U}, \nu; \mathbf{\mu})|^2 \right\} \right],
\end{align}
where $\sigma_{\rm rms}^2 = \sigma_N^2 / (N_B N_C)$ is the rms of the noise map. For uniform prior on parameters, $\Lambda(\mathbf{\mu})$ is nothing but the likelihood function.



\label{lastpage}
\end{document}